\documentclass[%
twocolumn,
groupedaddress,
amsmath,amssymb,
aps,
prb,
floatfix,superscriptaddress
]{revtex4-1}

\usepackage{mhchem}
\usepackage{amsmath}  
\usepackage{mathtools}
\usepackage{siunitx}
\usepackage{longtable}

\DeclareMathOperator*{\argmin}{argmin} 
\usepackage{epstopdf} 
\usepackage{graphicx}  
\usepackage{verbatim}  
\usepackage{color}   
\usepackage{subfigure} 
\usepackage{hyperref}  
\begin{document}

\title{
Anisotropic temperature-dependent lattice parameters and elastic constants from first principles}

\author{Samare~Rostami}
\email{samare.rostami@gmail.com}

\affiliation{European Theoretical Spectroscopy Facility, Institute of Condensed Matter and Nanosciences, Universit\'{e} catholique de Louvain, Chemin des \'{e}toiles 8, bte L07.03.01, B-1348 Louvain-la-Neuve, Belgium}

\author{Matteo~Giantomassi}
\affiliation{European Theoretical Spectroscopy Facility, Institute of Condensed Matter and Nanosciences, Universit\'{e} catholique de Louvain, Chemin des \'{e}toiles 8, bte L07.03.01, B-1348 Louvain-la-Neuve, Belgium}

\author{Xavier~Gonze}
\affiliation{European Theoretical Spectroscopy Facility, Institute of Condensed Matter and Nanosciences, Universit\'{e} catholique de Louvain, Chemin des \'{e}toiles 8, bte L07.03.01, B-1348 Louvain-la-Neuve, Belgium}

\begin{abstract}

The Quasi-harmonic Approximation (QHA) is a widely used method for calculating the temperature dependence of lattice parameters and the thermal expansion coefficients from first principles. 
However, applying QHA to anisotropic systems typically requires several dozens or even hundreds of phonon band structure calculations, leading to high computational costs. 
The Zero Static Internal Stress Approximation (ZSISA) QHA method partly addresses such caveat, but the computational load of its implementation remains high, so that its volumetric-only counterpart v-ZSISA-QHA is preferred. 
In this work, we present an efficient implementation of the ZSISA-QHA, enabling its application across a wide range of crystal structures under varying temperature (T) and pressure (P) conditions. 
By incorporating second-order derivatives of the vibrational free energy with respect to lattice degrees of freedom, we significantly reduce the number of required phonon band structure calculations for the determination of all lattice parameters and angles. For hexagonal, trigonal, and tetragonal systems, only six phonon band structure calculations are needed, while 10, 15, and 28 calculations suffice for orthorhombic, monoclinic, and triclinic systems, respectively. 
This method is tested for a variety of non-cubic materials, from uniaxial ones like ZnO and CaCO$_3$ to monoclinic or triclinic materials such as ZrO$_2$, HfO$_2$, and Al$_2$SiO$_5$, demonstrating a significant reduction in computational effort while maintaining accuracy in modeling anisotropic thermal expansion, unlike the v-ZSISA-QHA.
The method is also applied to the first-principles calculation of temperature-dependent elastic constants, with only up to six more phonon band structure calculations, depending on the crystallographic system.
\end{abstract}

\maketitle

\section{Introduction\label{sec:introduction}}
Understanding thermal expansion and elastic properties at varying temperatures and 
pressures is essential for predicting the thermomechanical behavior of crystalline 
materials in diverse applications. 
Except for the cubic crystallographic system, thermal expansion is inherently anisotropic, 
meaning it differs along various lattice directions. 
For the monoclinic and triclinic systems, the angle(s) will also change with temperature.
This directional dependence
plays a crucial role in material performance and the design of advanced devices. 
In certain cases, expansion may be negative or nearly zero along one direction while 
remaining positive along others, a key factor in controlling thermal stability and 
phase transitions~\cite{Barrera2005,Karunarathne2021,LEE2025}. First-principles methods, particularly those based on density 
functional theory (DFT)~\cite{Hohenberg1964, Kohn1965}, have been extensively used to compute volumetric thermal 
expansion. However, accurately capturing anisotropic thermal expansion remains 
computationally demanding, as it requires evaluating free energy derivatives along 
multiple independent lattice directions.

The quasiharmonic approximation (QHA)~\cite{Dove1993,Lazzeri1998,Carrier2007,Allen2020} is 
a well-established approach for modeling temperature-dependent material properties in solids 
with weak anharmonicity. This method accounts for changes in phonon frequencies due to lattice 
expansion while neglecting direct phonon-phonon interactions. QHA assumes that phonons remain 
harmonic, non-interacting, and primarily governed by lattice parameters and equilibrium atomic 
positions. 
Within this framework, the total free energy, including harmonic phonon contributions,
is expressed as a function of both lattice parameters and internal atomic positions at a given 
temperature. 
Minimizing this free energy for different temperatures provides insights into how these structural degrees of freedom evolve with temperature and pressure.

QHA has been widely used to predict thermal expansion by tracking the temperature-dependent 
evolution of lattice parameters and to compute elastic 
constants~\cite{Mathis2022,Mathis2024,Gong2024} through the evaluation of free
energy under applied strains. The approach enables the study of temperature-dependent mechanical 
properties and anisotropic lattice responses, making it a powerful tool for exploring 
thermoelastic behavior in a wide range of materials. The QHA Gibbs free energy consists of the phonon 
contribution, the Born-Oppenheimer (BO) energy at zero temperature, and the enthalpic term
associated with external pressure. For metals, an additional 
correction accounts for the
electronic free energy. Notably, even at absolute zero, the phonon free energy remains nonzero 
due to quantum zero-point motion.

Recent advances in first-principles techniques have enabled highly accurate
phonon spectrum calculations, facilitated by methods such as density functional 
perturbation theory (DFPT)~\cite{Baroni1987, Gonze1997, Baroni2001, Gonze2005a} and the
finite displacement method~\cite{Togo2015}. 
The reliability of these techniques 
has made high-throughput phonon calculations feasible, establishing QHA as 
a robust tool for predicting temperature-dependent thermodynamic and mechanical 
properties, provided that higher-order anharmonic effects remain minimal~\cite{Masuki2022,Allen2015,Masuki2022a}.

The computational cost of phonon spectra calculations remains a key limitation of 
QHA, making it significantly more demanding than the direct minimization of the
BO energy to determine equilibrium lattice parameters and atomic positions. This 
challenge is particularly relevant for materials with low symmetry, where thermal 
expansion depends on multiple independent structural degrees of freedom. In contrast, 
for cubic systems, where volume is the only free parameter and all internal atomic 
positions are fixed by symmetry, QHA simplifies to a one-dimensional optimization 
problem requiring relatively few phonon calculations.  
For lower-symmetry structures, including tetragonal, rhombohedral, hexagonal, 
orthorhombic, monoclinic, and triclinic crystals, thermal expansion becomes more 
complex as multiple lattice parameters evolve independently with temperature. 
Additionally, internal atomic positions, which are not fully constrained by symmetry, 
must be determined. Precisely accounting for these effects within QHA poses significant computational challenges and frequently requires further approximations.

One widely adopted approach is the zero static internal stress approximation (ZSISA), introduced by Allan and colleagues in 1996~\cite{Allan1996,Taylor1997,Taylor1999,Allen2020}.
ZSISA reduces computational complexity by assuming that internal atomic positions can be determined solely by minimizing the BO energy at each fixed lattice configuration, rather than performing a full free energy minimization. By enforcing the condition that atomic forces vanish in the BO energy landscape, ZSISA introduces only second-order errors in neglected thermal internal stresses. While this approximation is effective for predicting macroscopic thermal expansion, its accuracy in describing temperature-dependent internal atomic displacements is more limited, particularly at high temperatures or in systems where zero-point energy effects are significant~\cite{Liu2018,Masuki2023}. Despite these limitations, ZSISA remains a 
widely used approach for modeling anisotropic thermal expansion in QHA studies, 
particularly in uniaxial systems such as rhombohedral and wurtzite structures, where 
only two lattice degrees of freedom need to be considered~\cite{Lichtenstein2000, Mounet2005,Carrier2007a,Palumbo2017,Liu2018a,Ritz2018,Ritz2019,Li2021a,Brousseau2022,Gong2024}.  

For low-symmetry crystals such as orthorhombic, monoclinic, and triclinic structures,
the application of the ZSISA model becomes impractical due to the exponentially 
increasing computational cost of phonon spectrum calculations.
 Even in uniaxial systems, determining the temperature and pressure dependence of elastic constants and thermal expansion at high temperatures and pressures using ZSISA 
 remains computationally very expensive~\cite{Gong2024}. This is because, to accurately 
 capture high-pressure effects, the lattice undergoes significant changes, requiring 
 multiple QHA calculations for different volumes. 
Consequently, researchers often resort to the volume-constrained zero strain internal structure 
approximation (v-ZSISA)~\cite{Skelton2015}  to manage these computational challenges. In this approach, 
the volume is treated as the primary degree of freedom, while all other structural 
parameters are optimized at a fixed volume.

Within the v-ZSISA-QHA framework, phonon spectra are typically computed for only 
seven to twelve different volumes~\cite{Nath2016}, significantly fewer than in a standard ZSISA 
approach, where free energy sampling across all degrees of freedom leads to an 
exponential increase in computational effort.
Due to its reduced computational cost, most QHA studies in the literature rely on this 
approximation~\cite{Togo2010,Otero-de-la-Roza2011,Otero-de-la-Roza2011a,Li2011,Gupta2013,Togo2015,Skelton2015,Nath2016,Abraham2018}. 
However, while v-ZSISA-QHA provides reasonable predictions for the 
temperature and pressure dependence of volumetric expansion, its accuracy in capturing 
anisotropic thermal expansion is considerably limited, often failing to align with 
experimental results~\cite{Masuki2023}.

Various approximations have been introduced to reduce the computational cost associated 
with QHA. One such approach is the linear Gr{\"u}neisen  method~\cite{Allen2020}, which simplifies the problem 
by expanding the Born-Oppenheimer energy to second order and the phonon free energy to 
first order in the parameters being optimized. Unlike ZSISA and v-ZSISA, which reduce the 
effective dimensionality of the problem, the linear Gr{\"u}neisen  method maintains the full 
parameter space but achieves a more manageable scaling. This method estimates the 
derivative of the phonon free energy with respect to geometric parameters, such as volume,
using Gr{\"u}neisen parameters~\cite{Gruneisen1912}, requiring only a limited number of phonon calculations, 
typically twice the number of parameters considered. While effective for predicting 
zero-point lattice expansion and its contribution to zero-point renormalization  of the band gap energy in cubic 
and hexagonal materials~\cite{Brousseau2022}, the approach is inherently limited to the low-temperature regime 
(below the Debye temperature). At higher temperatures, its thermal expansion predictions 
become inaccurate, asymptotically saturating rather than continuing to increase as in QHA. 

In our previous work~\cite{Rostami2024}, we introduced new intermediate methods bridging
v-ZSISA and the linear Gr{\"u}neisen approach by applying a Taylor expansion to the 
vibrational free energy while keeping the BO energy calculations exact. 
Since BO energy computations are significantly cheaper than phonon calculations, this 
strategy reduces computational cost without introducing unnecessary approximations. 
 The most straightforward improvement over the linear
Gr{\"u}neisen method is to retain the full BO energy while using a minimal-order
Taylor expansion for the phonon free energy.

We proposed three variants: v-ZSISA-$\textrm{E}{\infty} \textrm{Vib}1$, v-ZSISA-$\textrm{E}{\infty} \textrm{Vib}2$, and v-ZSISA-$\textrm{E}{\infty} \textrm{Vib}4$, which employ first-, second-, and fourth-order Taylor expansions of the phonon free energy, requiring only two, three, and five phonon calculations at different volumes, respectively.

To evaluate their accuracy, we applied these methods to 12 materials spanning diverse space groups, from cubic to monoclinic structures. Using v-ZSISA-QHA as the reference, we found that a quadratic expansion (using three phonon calculations) yields highly accurate results with an error below 1\%, for our tested materials below 800 K, while a fourth-order expansion (five phonon calculations) closely matches the v-ZSISA-QHA reference. This highlights the effectiveness of our approach in significantly reducing computational costs while preserving high precision in volumetric thermal expansion predictions.

Accordingly, in the present work, we adopt a second-order Taylor expansion as a reliable approximation for the vibrational 
free energy. We extend the $\textrm{E}{\infty} \textrm{Vib}2$ method to 
anisotropic systems and introduce the ZSISA-$\textrm{E}{\infty} \textrm{Vib}2$ approach. 
This method applies a multidimensional Taylor expansion to the vibrational free energy, incorporating the system lattice degrees of freedom. By reducing the number of required
phonon calculations compared to a full multi-dimensional mesh, it maintains accuracy while 
significantly lowering computational costs. The approach determines thermal stress using 
finite differences and optimizes the Born-Oppenheimer energy self-consistently to eliminate thermal stress 
at a given pressure. Once the equilibrium structure is identified at each temperature and 
pressure, second derivatives of the BO energy, obtained through DFPT, and the vibrational free energy, computed using finite differences, allow for the calculation of thermal expansion coefficients and elastic constants. 

The validity of the proposed method is assessed by applying it to determine the 
anisotropic thermal expansion and elastic constants of materials with different 
crystallographic symmetries. The study includes cubic MgO, hexagonal ZnO, AlN, and GaN, trigonal \ce{CaCO3} and \ce{Al2O3}, tetragonal \ce{SnO2}, orthorhombic \ce{YAlO3}, monoclinic \ce{ZrO2}, \ce{HfO2}, and \ce{MgP4}, and triclinic \ce{Al2SiO5}.
For the uniaxial cases, our method exhibits excellent agreement with ZSISA-QHA. 
For crystals with even lower symmetries, a direct comparison is not performed with ZSISA-QHA, as it is too expensive. 
However, our predicted anisotropic thermal expansion aligns well 
with experimental data for such cases, whereas v-ZSISA fails to capture these trends. 
In some instances,
v-ZSISA even predicts an opposite trend, further emphasizing the advantage of our method 
in describing anisotropic effects.

The structure of this paper is as follows,
Sec.~\ref{sec:method} details the methodology, covering the definition of free energy, the 
quasiharmonic approximation (QHA), ZSISA and v-ZSISA, approximations for the vibrational 
free energy, thermal stress, determination of lattice parameters at finite temperature and 
external pressure, thermal expansion, and elastic constants. It also presents the 
equations for specific crystallographic cases, including orthorhombic and monoclinic 
structures.
Sec.~\ref{sec:Computational} describes the computational details and the materials studied.
Sec.~\ref{sec:result} presents our results, followed by the conclusions in Sec.~\ref{sec:conclusion}
Sec.~\ref{sec:appendix} provides an appendix with equations for the different crystal systems.

\section {method}\label{sec:method}

\subsection{The free energy }
Consider the crystallographic parameters, including lattice constants,
 cell angles, and internal atomic positions, represented by
  $C_\gamma$, with $\gamma$ ranging from $1$ to $N_\textrm{C}$. 
  In a scenario where symmetries are ignored, $N_\textrm{C}$ is given
 by $6 + 3N_{\textrm{at}} - 3$. Here, 6 represents
 the macroscopic crystallographic parameters, and $N_{\textrm{at}}$
 stands for the number of atoms within the primitive cell. The
 remaining $3N_{\textrm{at}} - 3$ parameters arise from the
 internal degrees of freedom, excluding the overall translational
motion of the crystal. However, symmetry considerations reduce
the number of truly independent crystallographic parameters
significantly. Additionally, this framework can be expanded to
include magnetic variables as part of the crystallographic
parameters, utilizing techniques such as
constrained-DFT~\cite{Dederichs1984,Gonze2022e}. 
The entire set of parameters $C_\gamma$ can be summarized as a
vector $\underline{C}$, and the temperature-dependent behavior
of these parameters, $C_\gamma(T)$ or $\underline{C}(T)$, is the focus of this study.

Although lattice constants and angles are well-defined macroscopic quantities, internal atomic positions represent 
average values over a large ensemble of cells that make up the solid. In this context, it is assumed that atomic 
position fluctuations within each cell occur around a single average value, excluding cases where multiple local configurations with similar (or nearly similar) energies are present, causing the system to transition between these 
configurations over time. Additionally, it is assumed that these crystallographic parameters can be continuously 
altered through the application of external stresses and internal forces within a computational framework, where the 
latter are applied uniformly across a sublattice associated with a specific average atomic position.

To determine the  temperature-dependent crystallographic parameters $C_\gamma(T)$ at zero pressure, the Helmholtz free energy $F[C_\gamma, T]$ must be minimized 
according to the following: 
\begin{align}
	 F(T) &=\min_{{C_\gamma}} [F(C_\gamma,T)], \label{eq:1}\\\
  C_\gamma(T) &= \argmin_{{C_\gamma}} [F(C_\gamma,T)]. \label{eq:2}\
	\end{align}
The temperature dependence of the parameters $C_\gamma(T)$ is therefore determined implicitly through the minimization condition 
given in Eq. (\ref{eq:1}), which leads to: 
\begin{align} 
\left.\frac{\partial F}{\partial C_\gamma} \right|_{\underline{C}(T)} = 0. \label{eq:3} 
\end{align}
The free energy, $F(\underline{C},T)$, is composed of several components: the Born-Oppenheimer
internal energy at absolute zero (0 K), which is independent of temperature; the vibrational (phonon)
contribution to the free energy; and additional corrections such as electronic entropy and
interactions between electrons and phonons. These corrections might be relevant for metals 
at very low temperatures but are generally negligible for insulators. Therefore, the  Helmholtz free energy can be approximated as:
\begin{align}
F(\underline{C},T) = E_{\textrm{BO}}(\underline{C}) + F_{\textrm{vib}}(\underline{C},T). \label{eq:4}
\end{align}

In this study, we assume that the BO energy, $E_{\textrm{BO}}(C_\gamma)$, can be computed quickly 
from first principles, especially compared to the time required to calculate the vibrational free energy, 
$F_{\textrm{vib}}(C_\gamma,T)$, which is also derived from first principles. However, it is important
to note that the gradients of $F_{\textrm{vib}}(C_\gamma,T)$ with respect to the parameters $C_\gamma$ 
are not directly available. They can, however, be estimated using finite difference methods. Each calculation of 
$F_{\textrm{vib}}(C_\gamma,T)$ for a different set of $C_\gamma$ needs to be meticulously planned.

By substituting Eq. (\ref{eq:4}) into Eq. (\ref{eq:3}), we obtain a more explicit condition for 
determining $\underline{C}(T)$:
\begin{align} -\frac{\partial E_{\textrm{BO}}}{\partial C_{\gamma}}\Big|_{\underline{C}(T)} = \frac{\partial F{\textrm{vib}}}{\partial C_{\gamma}}\Big|_{\underline{C}(T)}. \label{eq:5} \end{align}
The right side of Eq. (\ref{eq:5}) is referred to as the thermal gradient at the point $\underline{C}(T)$. If $\gamma$ 
corresponds to a cell parameter, then $\partial F_{\textrm{vib}}/\partial C_{\gamma}$ represents a stress. If $\gamma$ 
corresponds to an atomic position, it represents a generalized or collective force.

Since $F$ reaches a minimum at $C_{\gamma}(T)$, it follows from Eq.(\ref{eq:3}) that, for any $\gamma$ and $T$,
\begin{align}
\frac{\partial F}{\partial C_{\gamma}}\Big|_{C_{\gamma}(T),T} = 0.
\end{align}
Applying the chain rule to obtain the total derivative of this expression gives
\begin{align}
0 = \sum_{\gamma'} \frac{\partial^2 F}{\partial C_{\gamma} \partial C_{\gamma'}} \Big|_{C_{\gamma}(T),T} \frac{dC_{\gamma'}}{dT}\Big|_{T} + \frac{\partial^2 F}{\partial C_{\gamma} \partial T} \Big|_{C_{\gamma}(T),T}.
\end{align}
Consequently, $ \frac{dC_{\gamma}}{dT}\Big|_{T}$ can be obtained using the inverse of the second derivative matrix of the free energy, expressed as follows
\begin{align}
	\frac{dC_{\gamma}}{dT}\Big|_{T}=&-\sum_{\gamma'} 
	\left[\frac{\partial^2 F}{\partial C \partial C'} \Bigg|_{\underline{C}(T),T} \right]^{-1}_{\gamma \gamma'}
	\frac{\partial^2 F}{\partial C_{ \gamma'} \partial T } \Big|_{\underline{C}(T),T}\nonumber\\
	=&\sum_{\gamma'} 
	\left[\frac{\partial^2 F}{\partial C \partial C'} \Bigg|_{\underline{C}(T),T} \right]^{-1}_{\gamma \gamma'}
	\frac{\partial S}{\partial C_{ \gamma'} } \Big|_{\underline{C}(T),T},
 \label{eq:8}
\end{align}
where $S = -\frac{\partial F}{\partial T}$ represents the entropy.

\subsection{The Quasi-Harmonic Approximation (QHA)}

In the Quasi-Harmonic Approximation,
we treat atomic vibrations as harmonic, but we allow the vibrational 
frequencies to change depending on the crystallographic parameters and the positions of atoms within the structure.
To make this dependency clear, we express the frequencies as $\omega_{\textbf{q}\nu}(\underline{C})$, where
$\textbf{q}$ represents the phonon wavevector with $\nu$ the phonon branch index.
Such calculations are well-defined within a first-principles approach: the interatomic force constants 
are derived from the second-order derivatives of the BO energy, supposing that the 
$\underline{C}(T)$ parameters are those for which Eq.(\ref{eq:5})
is fulfilled at that temperature, namely, supposing that the BO gradient is cancelled by the thermal gradient.

The vibrational free energy is then calculated using Bose-Einstein statistics, 
$n_{\textbf{q}\nu}(\underline{C},T)$, which gives the occupation 
number for each phonon mode.  
This also includes the contribution
from zero-point motion. Specifically, the vibrational free energy per unit cell is

\begin{align} F_{\textrm{vib}}(\underline{C},T)&= \frac{1}{\Omega_{\textrm{BZ}}} \int_{\textrm{BZ}} \sum_{\nu} \nonumber\ \\&\Big( \frac{\hbar\omega_{\textbf{q}\nu}(\underline{C})}{2}
	k_\textrm{B}T \ln \big(1- e^{\frac{\hbar \omega_{\textbf{q}\nu}(\underline{C})}{k_\textrm{B}T}}\big) \Big) d\textbf{q}, \label{eq:8_} 
 \end{align}
and the entropy per unit cell is given by
\begin{align}
 S_{\textrm{vib}}(\underline{C},T)&=
 -\frac{dF_{\textrm{vib}}}{dT}\Big|_{\underline{C}}=
 \frac{k_\textrm{B}}{\Omega_{\textrm{BZ}}}
 \int_{\textrm{BZ}}
\sum_{\nu} 
\nonumber\\
&\Big(- 
\ln
\big(1-
e^{\frac{\hbar \omega_{\textbf{q}\nu}(\underline{C})}{k_\textrm{B}T}}\big)
+ n_{\textbf{q}\nu}\frac{\hbar \omega_{\textbf{q}\nu}(\underline{C})}{k_\textrm{B}T}
\Big)
d\textbf{q}
 \label{eq:9_}
\end{align}
where, $\Omega_{\textrm{BZ}}$ represents the Brillouin zone volume, related to the primitive cell
volume $\Omega_0$ by $\Omega_{\textrm{BZ}}=\frac{(2\pi)^3}{\Omega_0}$. It is important to note that
the phonon frequencies $\omega_{\textbf{q}\nu}$ do not depend directly on temperature, in contrast to  the phonon 
occupation numbers $n_{\textbf{q}\nu}$ that are given by Bose-Einstein statistics

\begin{align} n_{\textbf{q}\nu}(\underline{C},T)= \frac{1}{e^{\frac{\hbar \omega_{\textbf{q}\nu}(\underline{C})}{k_\textrm{B}T}}-1}. \label{eq:7_} \end{align}

\subsection{ZSISA and v-ZSISA }
In the ZSISA method, crystallographic parameters can be divided into external and internal strains. 
The internal strains are those associated with the internal degrees of freedom, such as atomic positions
within the unit cell, while the external strains refer to parameters like lattice constants, angles, or volume.
The approximation involves optimizing the internal strains by adjusting atomic positions, while the
external strains, such as lattice constants, are held constant.
This approach simplifies the problem by considering internal strains as a function of external strains. 

If the external strain is limited to changes in volume only, this approach is known as the v-ZSISA approximation. 
In this approach, the free energy defined in Eq.\ref{eq:4} is minimized at a fixed volume to determine the values 
of the remaining degrees of freedom, such as lattice parameters, angles, and internal atomic positions.

At a given temperature $T$ and equilibrium volume $V(T)$, the derivative of the free energy with respect to volume,
which corresponds to the pressure, is zero. This condition can be expressed as

\begin{align}
	 0&=P(V(T))= -\frac{\partial F} {\partial V}\Big|_{V(T),T}\\ \nonumber
	&= -\frac{\partial E_{\textrm{BO}}}{\partial V}\Big|_{V(T)}- \frac{\partial F_{\textrm{vib}}} {\partial V}\Big|_{V(T),T} \nonumber \\ &=P_{\textrm{BO}}(V(T))+P_{\textrm{vib}}(V(T),T), \label{eq:zero_pressure} 
\end{align}
where the $P_{\textrm{BO}}$ and $P_{\textrm{vib}}$ denote Born-Oppenheimer pressure  and the vibrational (thermal) pressure, respectively. At equilibrium volume for a specific temperature, these two pressures must cancel each other out:
\begin{align}
P_{\textrm{BO}}(V(T))=-P_{\textrm{vib}}(V(T),T).
\end{align}

In more general scenarios, where ZSISA is applied to external strains affecting lattice 
constants or angles, all components of the stress tensor must vanish. This condition is expressed as 
\begin{align}
	0=&\sigma_{ij}\left([R(T)]\right)=\frac{1}{V(T)}\frac{\partial F} {\partial \varepsilon_{ij}}\Big|_{[R(T)],T}\\
	=& \frac{1}{V(T)} \frac{\partial E_{\textrm{BO}}}{\partial \varepsilon_{ij}}\Big|_{[R(T)]}
+ \frac{1}{V(T)}\frac{\partial F_{\textrm{vib}}} {\partial \varepsilon_{ij}}\Big|_{[R(T)],T} \nonumber \\
=& \sigma_{ij}^{\textrm{BO}}({[R(T)]})+\sigma_{ij}^{\textrm{vib}}({[R(T)],T}),
\end{align}
 
where $\sigma_{ij}$ and $\varepsilon_{ij}$ are the $ij$-th components of the stress and strain tensors, respectively. 
$\sigma^{\textrm{BO}}$ and $\sigma^{\textrm{vib}}$ represent 
BO and vibrational (thermal) stresses. 
The set of lattice vectors $[R(T)]$ is defined as
 \begin{align}
 [R(T)]=	\left[ \begin{array}{lll}
 	\vec{R}_1 & 
 	\vec{R}_2& 
 	\vec{R}_3
 \end{array} \right]=
	\left[ \begin{array}{lll}
	R_{1,x} & R_{2,x}&R_{3,x}\\
    R_{1,y} & R_{2,y}&R_{3,y}\\
    R_{1,z} & R_{2,z}&R_{3,z} 
\end{array} \right].
 \end{align}
The temperature dependence of the vectors and their components have been omitted for the sake of compactness. Thus, the volume $V(T)$ is simply obtained as 
 \begin{align}
 V =&  | \det [R] | =|\vec{R}_1 \cdot (\vec{R}_2 \times \vec{R}_3)|
 \\ \nonumber
=& | R_{1,x} \left( R_{2,y} R_{3,z} - R_{2,z} R_{3,y} \right) - R_{1,y} \left( R_{2,x} R_{3,z} 
 \right.\\ \nonumber
&\left .  - R_{2,z} R_{3,x} \right) + R_{1,z} \left( R_{2,x} R_{3,y} - R_{2,y} R_{3,x} \right) | .
\end{align}

Subsequently, at equilibrium, the BO and vibrational stresses must cancel each other out:
\begin{align}
 \sigma_{ij}^{\textrm{BO}}({[R(T)]})=-\sigma_{ij}^{\textrm{vib}}({[R(T)],T}).
\end{align}
When considering a constant external pressure
 $P_{\textrm{ext}}$, the Gibbs free energy 
  $G([R(T)],T)$ is given by
 \begin{align} 
 &G([R(T,P_{\textrm{ext}})],T,P_{\textrm{ext}}) = E_{\textrm{BO}}([R(T,P_{\textrm{ext}})]) \\
&\qquad\qquad + 
 F_{\textrm{vib}}([R(T,P_{\textrm{ext}})],T) + P_{\textrm{ext}}V(T,P_{\textrm{ext}}). \nonumber\label{eq:Gibbs1}
  \end{align}
Thus, the derivative of the Gibbs free energy with respect to volume in v-ZSISA becomes
\begin{align}
	0&= -\frac{\partial G} {\partial V}\Big|_{V(T,P_{\textrm{ext}}),T,P_{\textrm{ext}}}\\ \nonumber
	&= -\frac{\partial E_{\textrm{BO}}}{\partial V}\Big|_{V(T,P_{\textrm{ext}})}- \frac{\partial F_{\textrm{vib}}} {\partial V}\Big|_{V(T,P_{\textrm{ext}}),T}+P_{\textrm{ext}} 
\end{align}
which simplifies to:
\begin{align}
P_{\textrm{BO}}(V(T,P_{\textrm{ext}}))=-P_{\textrm{vib}}(V(T,P_{\textrm{ext}}),T)-P_{\textrm{ext}}.\label{eq:zero_pressure_condition} 
\end{align}
Similarly, for ZSISA, if a uniform external pressure is applied, it will affect the 
diagonal components of the stress tensor. Specifically, the total stress tensor
$\sigma_{ij}$ under a uniform external pressure $P_{\textrm{ext}}$ is modified by this external pressure and is such that:
\begin{align} 
\sigma_{ii}^{\textrm{BO}}([R(T,P_{\textrm{ext}})])= -\sigma_{ii}^{\textrm{vib}}([R(T,P_{\textrm{ext}})],T) - P_{\textrm{ext}}. \label{eq:18}\end{align}
In this context, the external pressure effectively adds a term to the diagonal
components of the stress tensor. 
Such modification results in a shift in the equilibrium conditions and can influence the response of the material to external stresses.
All this might be trivially generalized to the case of anisotropic external stress (not developed in this work, though).
 
In both v-ZSISA-QHA and ZSISA-QHA methods, the interpolation of free energies necessitates phonon spectra calculations
at multiple volumes, with the number of calculations determined by the degrees of freedom (DOF) of the lattice. 
This process can be computationally demanding.
For the case where only the volume is considered, or in the case of a cubic system with a single DOF for the lattice constant, at least five phonon spectra calculations are required to adequately fit an equation of state (EOS).

In systems with two DOFs, such as hexagonal, 
trigonal, or tetragonal lattices, a grid of 5 $\times$ 5 points 
(five for each DOF) is needed to interpolate a parabola to determine the free energy. As the number of DOFs increases, 
as in orthorhombic (3 DOFs), monoclinic (4 DOFs), and triclinic (6 DOFs) lattices, the number of necessary phonon 
calculations grows exponentially, scaling as $5^3$, $5^4$, and $5^6$, respectively. Although these calculations are 
significantly fewer compared to the full QHA, they still pose substantial computational 
challenges for many systems.

To address this issue, various approximations can be employed to achieve comparable accuracy while reducing the 
computational cost of  these calculations. In our previous work, we proposed approximations
for v-ZSISA-QHA that utilize a Taylor expansion of $F_{\textrm{vib}}$
to reduce the number of required phonon band structure calculations, while preserving the BO energies at multiple volumes. 
In that study, we examined the accuracy of the Taylor expansion approximation for the 
vibrational free energy and determined the necessary expansion order. Results indicated 
that, for most materials, including terms up to the second derivative, with three phonon band structure calculations, is sufficient to reproduce the results of the QHA. We referred to this approach as
${\textrm{E}{\infty} \textrm{Vib}2}$ approximation.
This finding provides a foundation for extending the advantages of 
${\textrm{E}{\infty} \textrm{Vib}2}$ to complex crystallographic systems with higher 
lattice degrees of freedom. In this work, we propose a new approximation method designed 
to accommodate systems with any number of degrees of freedom,  from uniaxial to 
triclinic crystals.

\subsection{Approximation of the vibrational free energy}

In ZSISA, the lattice vectors correspond to the external strains, and the internal atomic coordinates, which minimize the atomic forces, are functions of these strains.
To account for the vibrational contributions, 
the phonon free energy is expanded around a reference crystallographic lattice 
configuration $ [R^{\bullet}$] using a Taylor series in terms of the
strain deviation, stopping at second order.

According to the ${\textrm{E}{\infty} \textrm{Vib}2}$ approximation, 
to derive the free energy from Eq.~\ref{eq:Gibbs1}, the expression for $E_{\textrm{BO}}
([R(T)])$ is preserved without any approximations. To incorporate vibrational 
contributions, the phonon free energy $F_{\textrm{vib}}$ is expanded around a reference crystallographic 
lattice configuration using a Taylor series in terms of strain deviation, stopped at 
the second order.

To facilitate the analysis, we define the strain matrix $[\varepsilon^{\textrm{BO}}]$ as follows:
\begin{align}
   [\varepsilon^{\textrm{BO}}]= \left[ \begin{array}{lll}
			\varepsilon^{\textrm{BO}}_{xx} & \varepsilon^{\textrm{BO}}_{xy} & \varepsilon^{\textrm{BO}}_{xz} \\ 
			\varepsilon^{\textrm{BO}}_{yx} & \varepsilon^{\textrm{BO}}_{yy} & \varepsilon^{\textrm{BO}}_{yz} \\ 
			\varepsilon^{\textrm{BO}}_{zx} & \varepsilon^{\textrm{BO}}_{zy} & \varepsilon^{\textrm{BO}}_{zz}
		\end{array} \right].
\end{align}
Here, $\varepsilon^{\textrm{BO}}_{xx}$, $\varepsilon^{\textrm{BO}}_{yy}$, $\varepsilon^{\textrm{BO}}_{zz}$ are the normal strains along 
the $x$, $y$, and $z$ directions, respectively, while the off-diagonal terms
$\varepsilon^{\textrm{BO}}_{xy}$, $\varepsilon^{\textrm{BO}}_{xz}$, $\varepsilon^{\textrm{BO}}_{yz}$ represent shear strains.

Consider $[R^{\textrm{BO}}]$, the lattice vectors of the structure at the minimized BO energy. The deformed lattice vectors under applied strains can be defined as:
    	\begin{align} \label{eq:R}
		[R]&=([1]+[\varepsilon^{\textrm{BO}}]).[R^{BO}] \\
  &= 
			\left[ \begin{array}{lll}
			1+\varepsilon^{\textrm{BO}}_{xx} & \varepsilon^{\textrm{BO}}_{xy} & \varepsilon^{\textrm{BO}}_{xz} \\ 
			\varepsilon^{\textrm{BO}}_{yx} & 1+\varepsilon^{\textrm{BO}}_{yy} & \varepsilon^{\textrm{BO}}_{yz} \\ 
			\varepsilon^{\textrm{BO}}_{zx} & \varepsilon^{\textrm{BO}}_{zy} & 1+\varepsilon^{\textrm{BO}}_{zz}
		\end{array} \right]
        	\left[ \begin{array}{lll}
			R^{\textrm{BO}}_{1,x} & 	R^{\textrm{BO}}_{2,x} & 	R^{\textrm{BO}}_{3,x}  \\ 
			R^{\textrm{BO}}_{1,y} & 	R^{\textrm{BO}}_{2,y} & 	R^{\textrm{BO}}_{3,y}  \\
			R^{\textrm{BO}}_{1,z} & 	R^{\textrm{BO}}_{2,z} & 	R^{\textrm{BO}}_{3,z}  
		\end{array} \right].\nonumber
	\end{align} 
  This matrix transformation captures how the lattice vectors deform under the influence of strain. 
  This deformation results in:
 \begin{widetext}
  \begin{align}
		[R]&=	 \left[ \begin{array}{lll}
		(1+\varepsilon^{\textrm{BO}}_{xx}) R^{\textrm{BO}}_{1,x}+   \varepsilon^{\textrm{BO}}_{xy}   R^{\textrm{BO}}_{1,y}+   \varepsilon^{\textrm{BO}}_{xz}   R^{\textrm{BO}}_{1,z} 
& (1+\varepsilon^{\textrm{BO}}_{xx})  R^{\textrm{BO}}_{2,x}+   \varepsilon^{\textrm{BO}}_{xy}   R^{\textrm{BO}}_{2,y}+   \varepsilon^{\textrm{BO}}_{xz}   R^{\textrm{BO}}_{2,z} &  
(1+\varepsilon^{\textrm{BO}}_{xx}) R^{\textrm{BO}}_{3,x}+   \varepsilon^{\textrm{BO}}_{xy}  R^{\textrm{BO}}_{3,y}+   \varepsilon^{\textrm{BO}}_{xz}    R^{\textrm{BO}}_{3,z} \\
		 \varepsilon^{\textrm{BO}}_{yx}   R^{\textrm{BO}}_{1,x}+ (1+\varepsilon^{\textrm{BO}}_{yy}) R^{\textrm{BO}}_{1,y}+   \varepsilon^{\textrm{BO}}_{yz}   R^{\textrm{BO}}_{1,z} 
         &  \varepsilon^{\textrm{BO}}_{yx}   R^{\textrm{BO}}_{2,x}+ (1+\varepsilon^{\textrm{BO}}_{yy}) R^{\textrm{BO}}_{2,y}+   \varepsilon^{\textrm{BO}}_{yz}   R^{\textrm{BO}}_{2,z} 
        &  \varepsilon^{\textrm{BO}}_{yx}  R^{\textrm{BO}}_{3,x}+ (1+\varepsilon^{\textrm{BO}}_{yy}) R^{\textrm{BO}}_{3,y}+   \varepsilon^{\textrm{BO}}_{yz}   R^{\textrm{BO}}_{3,z} \\ 
		 \varepsilon^{\textrm{BO}}_{zx}   R^{\textrm{BO}}_{1,x}+   \varepsilon^{\textrm{BO}}_{zy}   R^{\textrm{BO}}_{1,y}+ (1+\varepsilon^{\textrm{BO}}_{zz})  R^{\textrm{BO}}_{1,z} 
		&  \varepsilon^{\textrm{BO}}_{zx}   R^{\textrm{BO}}_{2,x}+   \varepsilon^{\textrm{BO}}_{zy}   R^{\textrm{BO}}_{2,y}+ (1+\varepsilon^{\textrm{BO}}_{zz})  R^{\textrm{BO}}_{2,z}  	
		&  \varepsilon^{\textrm{BO}}_{zx}   R^{\textrm{BO}}_{3,x}+   \varepsilon^{\textrm{BO}}_{zy}   R^{\textrm{BO}}_{3,y}+ (1+\varepsilon^{\textrm{BO}}_{zz})  R^{\textrm{BO}}_{3,z} 
	\end{array} \right].
	\end{align} 
 \end{widetext}
In this formulation, the independent parameters are the strain components, allowing the equation to be expressed in terms of these strains. 

The phonon free energy is then expressed as a Taylor series expansion around a 
reference lattice configuration $[R^{\bullet}]$, that might be equal to the BO configuration or not, which is defined by the strain 
$[\varepsilon^{\textrm{BO} \bullet}]$. The expansion is written in terms of the 
strain deviation, $\Delta^{\bullet}[\varepsilon^{\textrm{BO}}] = [\varepsilon^{\textrm{BO}}] - [\varepsilon^{\textrm{BO}\bullet}]$, and is truncated at the second order:

\begin{align}
	F_{\textrm{vib}}[[R],T] &= F_{\textrm{vib}}[[R^{\bullet}],T]
    \nonumber
    \\ 
	+&\sum_{\gamma} \Delta ^{\bullet} \varepsilon^{\textrm{BO}}_{\gamma} . \frac{\partial F_{\textrm{vib}}}{\partial \varepsilon^{\textrm{BO}}_{\gamma}}\Big|_{[R^{\bullet}],T}
    \nonumber
    \\
	+&\frac{1}{2}\sum_{\gamma \gamma'} \Delta ^{\bullet} \varepsilon^{\textrm{BO}}_{\gamma} . \Delta ^{\bullet} \varepsilon^{\textrm{BO}}_{\gamma'} . \frac{ \partial ^2F_{\textrm{vib}}}{ \partial \varepsilon^{\textrm{BO}}_{\gamma} \partial \varepsilon^{\textrm{BO}}_{\gamma'}}\Big|_{[R^{\bullet}],T},
    \label{eq:22}
\end{align}
where $\gamma$ represents the indices \{$xx, yy, zz, xy, xz, yz$\}, corresponding 
to the non-zero components of $\varepsilon^{\textrm{BO}}_{\gamma}$.

With $n$ representing the number of lattice degrees of freedom, the minimum number of 
configurations required to compute the Taylor expansion is
$1 + 2n + {n(n-1)}/{2} = {(n+1)(n+2)}/{2}$.

To compute the diagonal second derivatives ${\partial^2 F_{\textrm{vib}}}/{\partial (\varepsilon^{\textrm{BO}}_{\gamma}})^2$ 
at the reference configuration $[\varepsilon^{\textrm{BO}\bullet}]$, three points are required for each value of $\gamma$: 
one at the central configuration $[\varepsilon^{\textrm{BO}\bullet}]$, and two additional 
configurations corresponding to deformations 
$([\varepsilon^{\textrm{BO}\bullet}]\pm \delta \varepsilon^{\textrm{BO}}_{\gamma})$.
This results in a total of $1 + 2n$ configurations.

For the mixed second derivative 
${\partial^2 F_{\textrm{vib}}}/{\partial \varepsilon^{\textrm{BO}}_{\gamma} \partial \varepsilon^{\textrm{BO}}_{\gamma'}}$,
four points are typically needed in a simple approach corresponding to
$([\varepsilon^{\textrm{BO}\bullet}] \pm \delta \varepsilon^{\textrm{BO}}_{\gamma} \pm \delta \varepsilon^{\textrm{BO}}_{\gamma'})$, which are
symmetrically distributed around $[\varepsilon^{\textrm{BO}\bullet}]$. Instead, to minimize the number 
of required calculations, we arrange the points as follows: 
$[\varepsilon^{\textrm{BO}\bullet}]$, $([\varepsilon^{\textrm{BO}\bullet}] - \delta \varepsilon^{\textrm{BO}}_{\gamma})$, 
$([\varepsilon^{\textrm{BO}\bullet}] - \delta \varepsilon^{\textrm{BO}}_{\gamma'})$, and $([\varepsilon^{\textrm{BO}\bullet}] - \delta \varepsilon^{\textrm{BO}}_{\gamma} - \delta \varepsilon^{\textrm{BO}}_{\gamma'})$.
Three of these points overlap with those used to compute
${\partial^2 F_{\textrm{vib}}}/{\partial {(\varepsilon^{\textrm{BO}}_{\gamma}})^2}$ and
${\partial^2 F_{\textrm{vib}}}/{\partial {(\varepsilon^{\textrm{BO}}_{\gamma'}})^2}$,
so only one additional configuration is required for each unique pair of 
$\gamma$ and $\gamma'$. This results in a total of ${n(n-1)}/{2}$ extra points.

The choice of the reference lattice configuration $[R^{\bullet}]$ plays an important role in ensuring accurate results. If $[R^{\bullet}]$ falls within the expected range of lattice parameters across different temperatures and pressures, the calculations yield more reliable predictions. In our previous work on the v-ZSISA approach, we demonstrated that for materials with positive thermal expansion, shifting the reference structure away from the BO lattice in the direction of expansion improves accuracy. A similar strategy is applied here, where $[R^{\bullet}]$   is adjusted based on the anticipated thermal expansion and pressure effects. When external pressure is applied, materials typically exhibit positive thermal expansion, yet their lattice parameters at zero temperature become smaller than the BO configuration at zero pressure. Therefore, selecting an appropriate $[R^{\bullet}]$ that accounts for both thermal and pressure-induced effects is crucial for achieving precise results.
In the present work, since we are considering low-pressure conditions, we focus primarily on the thermal expansion behavior. Accordingly, we introduce a positive shift in $[R^{\bullet}]$ to ensure that our calculations yield the expected positive thermal expansion of the studied materials. However, at high pressures, an alternative approach is to define the BO structure as the lattice relaxed at that pressure, rather than at zero pressure. In this case, applying a positive shift in the direction of thermal expansion remains an effective strategy for accurately capturing the material’s behavior.

\begin{figure}[t!]
\includegraphics[width=0.45\textwidth]{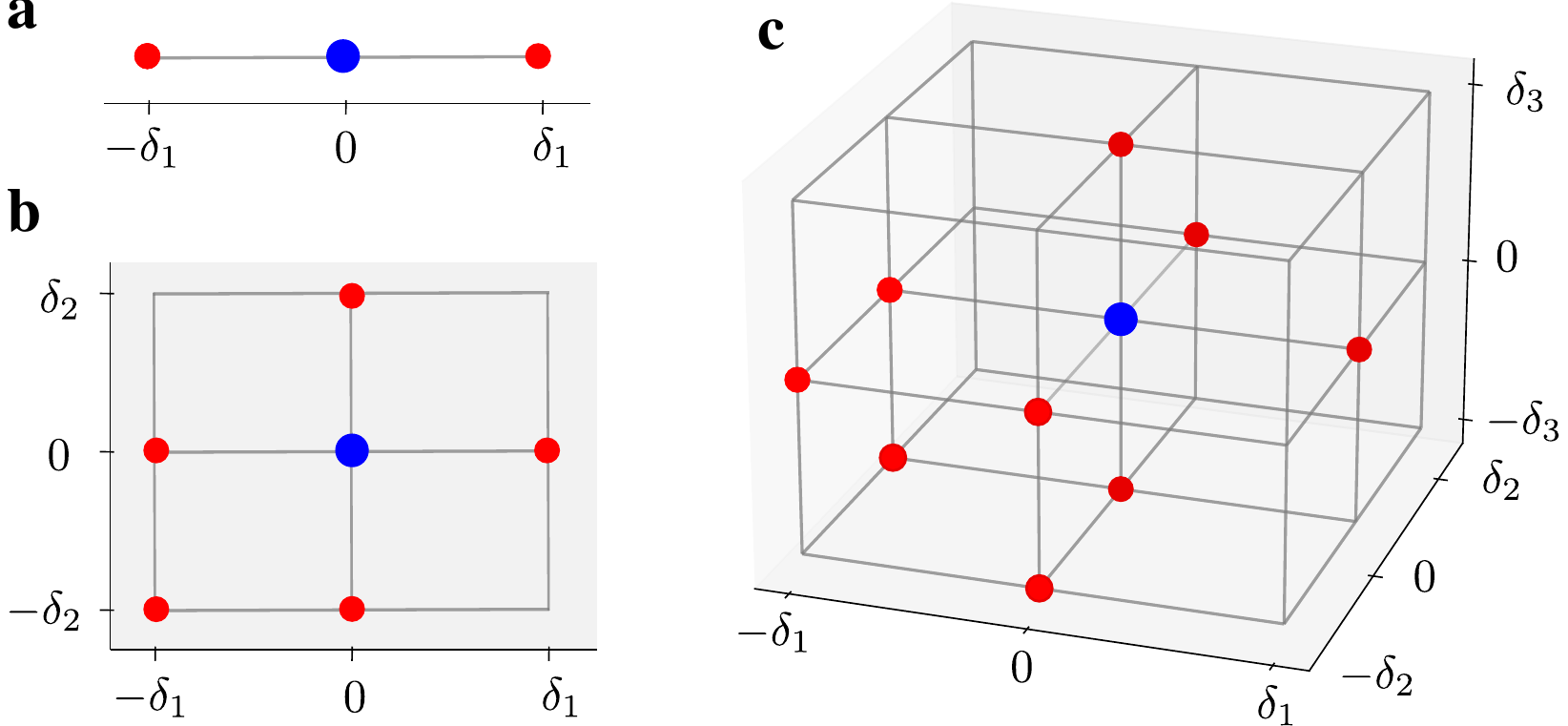}
  \caption{Representation of the points where phonon spectra calculations are performed, for configurations with (a) 1 DOF, 
  (b) 2 DOF, and (c) 3 DOF. The blue point denotes the center of calculations
  (indicated by $[\varepsilon^{\textrm{BO}\bullet}]$), while the red points represent the 
  deformations required for each degree of freedom 
  $\varepsilon^{\textrm{BO}}_{\gamma}$
  . The total
  number of phonon spectra calculations for 1, 2, and 3 DOF are 3, 6, and 10, respectively.}
	\label{fig:points}
\end{figure}

Figure~\ref{fig:points} 
presents the points where phonon spectra calculations are performed for 
systems with 1, 2, and 3 degrees of freedom.  
The central blue point is the reference calculation 
$[\varepsilon^{\textrm{BO}\bullet}]$, while red points indicate the
additional calculation needed to obtain finite difference data up to second degree for each degree of freedom $\varepsilon^{\textrm{BO}}_\gamma$, including cross derivatives. For 1, 2, and 3 DOF, the total number 
of calculations are 3, 6, and 10, respectively. Additionally, for systems with 
4 and 6 DOF, 15 and 28 calculations are required.
In Table~\ref{tab:I},  the number of degrees of freedom and
the corresponding required number of phonon spectra calculations for various crystal structures are listed.

At this stage, the vibrational contribution 
to
$F_{E_{\infty}\textrm{Vib}2}([R],T)$ from Eq.(\ref{eq:Gibbs1}), is obtained by quadratic extrapolation over the $n$-dimensional surface.
We are left with the BO contribution evaluation.
If we perform a fit of the $n$-dimensional BO energy  and determine the lattice parameters by minimizing the free energy, as expressed in Eq.(\ref{eq:2}), we face still a problem with the explosion of the number of such BO calculations.
For example, using five interpolation points along each DOF to obtain a multidimensional BO grid, leads to a total number of configurations scaling as $5^n$, which grows exponentially with the number of DOF. For each of these $5^n$ configurations, the BO energy must also satisfy the ZSISA condition, requiring full atomic position relaxation which is a computationally intensive process. In such an approach, one does not take benefit from the well-known easy computation of forces and stresses for the DFT BO calculation.
To overcome this challenge, we propose an alternative approach based on thermal stress, which bypasses the need to fit an $n$-dimensional surface.

\subsection{Thermal stress}

Once $F_{\textrm{vib}}$ is determined, Eq.(\ref{eq:22}), its derivatives can be evaluated directly as:
\begin{align}\label{eq:28_1}
	\frac{dF_{\textrm{vib}}}{d\varepsilon^{\textrm{BO}}_{\gamma}} \big|_{[R],T}= 
	\frac{\partial F_{\textrm{vib}}}{\partial \varepsilon^{\textrm{BO}}_{\gamma}}\Big|_{[R ^{\bullet}],T}
	+\sum_{ \gamma'}  \Delta ^{\bullet} \varepsilon^{\textrm{BO}}_{\gamma'} . \frac{ \partial ^2F_{\textrm{vib}}}{ \partial \varepsilon^{\textrm{BO}}_{\gamma} \partial \varepsilon^{\textrm{BO}}_{\gamma'}}\Big|_{[R ^{\bullet}],T}.
\end{align}
The thermal stress $\sigma_{\gamma}^{\textrm{vib}}({[R(T,P_{\textrm{ext}})],T})$ is then obtained by:
\begin{align}\label{eq:30}
	&\sigma_{\gamma}^{\textrm{vib}}({[R(T,P_{\textrm{ext}})],T,P_{\textrm{ext}}})=
    \nonumber
    \\
	~~~ &\frac{1}{V(T,P_{\textrm{ext}})}\sum_{ \gamma'}\frac{d F_{\textrm{vib}}} {d \varepsilon^{\textrm{BO}}_{\gamma'}}\Big|_{[R(T,P_{\textrm{ext}})],T} \left(\frac{d \varepsilon^{\textrm{BO}}_{\gamma'} } {d \varepsilon_{\gamma}} \right)\Big|_{[R(T,P_{\textrm{ext}})]}.
\end{align}
where $\varepsilon$ is the strain referenced specifically to the structure at which the derivative is performed, here, 
$[R(T,P_{\textrm{ext}})]$, rather than the reference structure $[R^{\textrm{BO}}]$ used for $\varepsilon^{\textrm{BO}}$.
The scaling factor $d\varepsilon^{\textrm{BO}} / d\varepsilon$ is necessary to ensure that strain is always derived from the specific structure under consideration. In Eq.~(\ref{eq:R}), $[R]$ is defined as a strain applied to $R^{\textrm{BO}}$. However, this definition is not unique and the strain can be defined with respect to other configurations as well. 
In particular, the stress definition is the derivative evaluated at a state where the strain is zero. Therefore, this scaling is essential to maintain consistency.

To obtain the conversion factor, we define the applied strain $[\varepsilon]$, by expressing $[R]$ with respect to $[R(T,P_\textrm{ext})]$:
\begin{align}
    [R]&=\big([1]+[\varepsilon]\big).[R(T,P_{\textrm{ext}})].
\end{align}
By substituting this relation into Eq.~(\ref{eq:R}), we arrive at the following expression:
\begin{align}
    [1]+[\varepsilon^{\textrm{BO}}]&=([1]+[\varepsilon]).[R(T,P_\textrm{ext})][R^{\textrm{BO}}]^{-1}.
    \end{align}
Differentiating with respect to $[\varepsilon]$, we obtain for all values of $k$:
    \begin{align}
    \frac{d \varepsilon_{kl}^{\textrm{BO}}}{d \varepsilon_{kj}}\Big|_{[R(T,P_\textrm{ext})]}&= \sum_{i}
    [R(T,P_\textrm{ext})]_{ ji}[R^{\textrm{BO}}]^{-1}_{il}.
\end{align}
In the special case where $[R]=[R(T,P_{\textrm{ext}})]$, the strain $[\varepsilon]$ becomes zero. Let $\varepsilon^{\textrm{BO}}(T,P_\textrm{ext})$ denote the strain corresponding to 
$[R(T,P_\textrm{ext})]$. 
Substituting this into the equation, the strain derivative simplifies to:
\begin{align}
  \frac{d \varepsilon_{kl}^{\textrm{BO}}}{d \varepsilon_{kj}}\Big|_{[R(T,P_{\textrm{ext}})]}
    &=[1]_{jl}+[\varepsilon^{\textrm{BO}}(T,P_\textrm{ext})]_{jl}.
\end{align}

Using these thermal stresses, the BO stress can be evaluated via Eq.~(\ref{eq:18}). Knowing the BO stress, the optimized BO configuration can be found by enforcing a constraint to maintain the correct stress. 

The optimization of a geometric structure in order to have zero net atomic force and zero stress is common in all electronic structure packages. 
It is usually also possible to optimize a geometric structure under a given external pressure. 
The relaxation under an arbitrary external anisotropic stress might not be so common. 
This is an advanced feature implemented in ABINIT for a long time. One can relax both the lattice and atomic positions with a specified external stress tensor and specified atomic forces, using the usual algorithms for geometric relaxation. 
This is quite efficient and well-tested, and the computational effort is much lower than the computation of a full phonon band structure.

\begin{table}
\caption{Non-zero components of the strain matrix $[\varepsilon]$ for various crystallographic structures. 
The symmetry of each crystal structure imposes constraints on the strain matrix, allowing 
only specific components to be non-zero. The number of independent degrees of freedom (DOF) 
in the lattice and the number of deformations (\#deform) required for phonon spectra calculations are also reported.}
\label{tab:I}

	\begin{ruledtabular}
		\begin{tabular}{lccc}
   Crystal & Strain & DOF & \#deform\\
  \hline
  \textbf{Bulk}\\
  Cubic & $\varepsilon_{xx}=\varepsilon_{yy}=\varepsilon_{zz}$ & 1 & 3 \\
  \hline
  \begin{tabular}{l}  Hexagonal \\ Trigonal\\  Tetragonal \end{tabular}
 & $\varepsilon_{xx}=\varepsilon_{yy}~ ,~\varepsilon_{zz}$ & 2 & 6\\
  \hline
Orthorhombic & $\varepsilon_{xx}~,~\varepsilon_{yy} ~,~\varepsilon_{zz}$  & 3 & 10\\
			\hline
			 Monoclinic  &  $\varepsilon_{xx}~,~\varepsilon_{yy} ~,~\varepsilon_{zz},~\varepsilon_{xz} $ & 4 & 15\\
    \hline
			Triclinic  &  \begin{tabular}{l} $\varepsilon_{xx}~,~\varepsilon_{yy} ~,~\varepsilon_{zz}$\\$\varepsilon_{xz}
   ~,~\varepsilon_{xy}  ~,~\varepsilon_{yz}$ \end{tabular} & 6 & 28\\
    \hline
  \textbf{Slab} \\
  Isotropic & $\varepsilon_{xx}=\varepsilon_{yy}$ & 1 & 3 \\
  \hline
  Anisotropic (2DOF) 
 & $\varepsilon_{xx}~ ,~\varepsilon_{yy}$ & 2 & 6\\
  \hline
Anisotropic (3DOF)  & $\varepsilon_{xx}~,~\varepsilon_{yy} ~,~\varepsilon_{xy}$  & 3 & 10
    \end{tabular}
	\end{ruledtabular}
\end{table}

Due to symmetry constraints in various crystallographic structures, certain components of the strain matrix $[\varepsilon]$ are inherently zero, meaning that not all stresses need to be computed. Table \ref{tab:I}
 lists the non-zero components of the strain matrix for different primitive cell structures, showing how these symmetry restrictions reduce the number of independent degrees of freedom in each configuration. In all crystals except monoclinic, triclinic,
 and anisotropic slab with 3 DOFs only normal (diagonal) strains are non-zero, with shear strains absent. For cubic crystals, all diagonal strains are equal, while in hexagonal, trigonal, and tetragonal crystals, two diagonal strains ($\varepsilon_{xx}$ and $\varepsilon_{yy}$) are equal, allowing $\varepsilon_{zz}$ to vary independently.
The number of deformations required to determine the stress components follows the 
same approach discussed in the previous section. 

After determining the thermal stress at a given temperature, the next step is to solve Eq.~(\ref{eq:18}) 
self-consistently to obtain the final lattice configuration and corresponding properties.

 \subsection{Finding lattice parameters at \(T\) and \(P_{\textrm{ext}}\)}

The process of finding the optimal lattice parameters at a specified temperature \(T\) and external 
pressure \(P_{\textrm{ext}}\) involves several iterative steps. We begin by generating deformations
from an initial configuration \(\varepsilon^{\textrm{BO}\bullet}\), applying strains consistent with the 
system symmetries. From these, we determine the vibrational free energy and thermal stress, which 
provide the basis for identifying the optimal volume and lattice parameters.

The first step in this process is to obtain an initial guess for the lattice configuration \([R]\). 
Using this guess, we compute the thermal stress. Since this \([R]\) does not necessarily correspond to 
the minimum free energy at temperature \(T\), the condition in Eq.~(\ref{eq:18}) is not satisfied initially.

To address this, we define a target stress as follows:

\begin{align}
\sigma_{ij}^{\textrm{target}}([R],T,P_{\textrm{ext}})= 
\begin{cases}
-\sigma_{ii}^{\textrm{vib}}([R], T) - P_{\textrm{ext}} & \quad \text{if } i = j \\
-\sigma_{ij}^{\textrm{vib}}([R], T) & \quad \text{if } i \neq j
\end{cases}
\end{align}

The value of $\sigma_{ij}^{\textrm{vib}}([R], T)$ is determined from the quadratic approximation, so without doing any recomputation of the phonon band structure.
The goal is to find a lattice configuration \([R(T,P_{\textrm{ext}})]\) such that the target stress \(\sigma_{ij}
^{\textrm{target}}([R(T,P_{\textrm{ext}})],T,P_{\textrm{ext}})\) matches the BO stress \(\sigma_{ij}^{\textrm{BO}}
([R(T)])\). While the BO stress may differ for the initial guess, we can solve this self-consistently.

Using our initial guess, we compute the target stress. With this target stress, we relax the lattice 
and atomic positions by minimizing forces while imposing the constraint that the target stress is 
achieved, following the ZSISA approach. This relaxation step alters the lattice until the stress target is met.
\begin{figure}[t!]
		\includegraphics[width=0.45\textwidth]{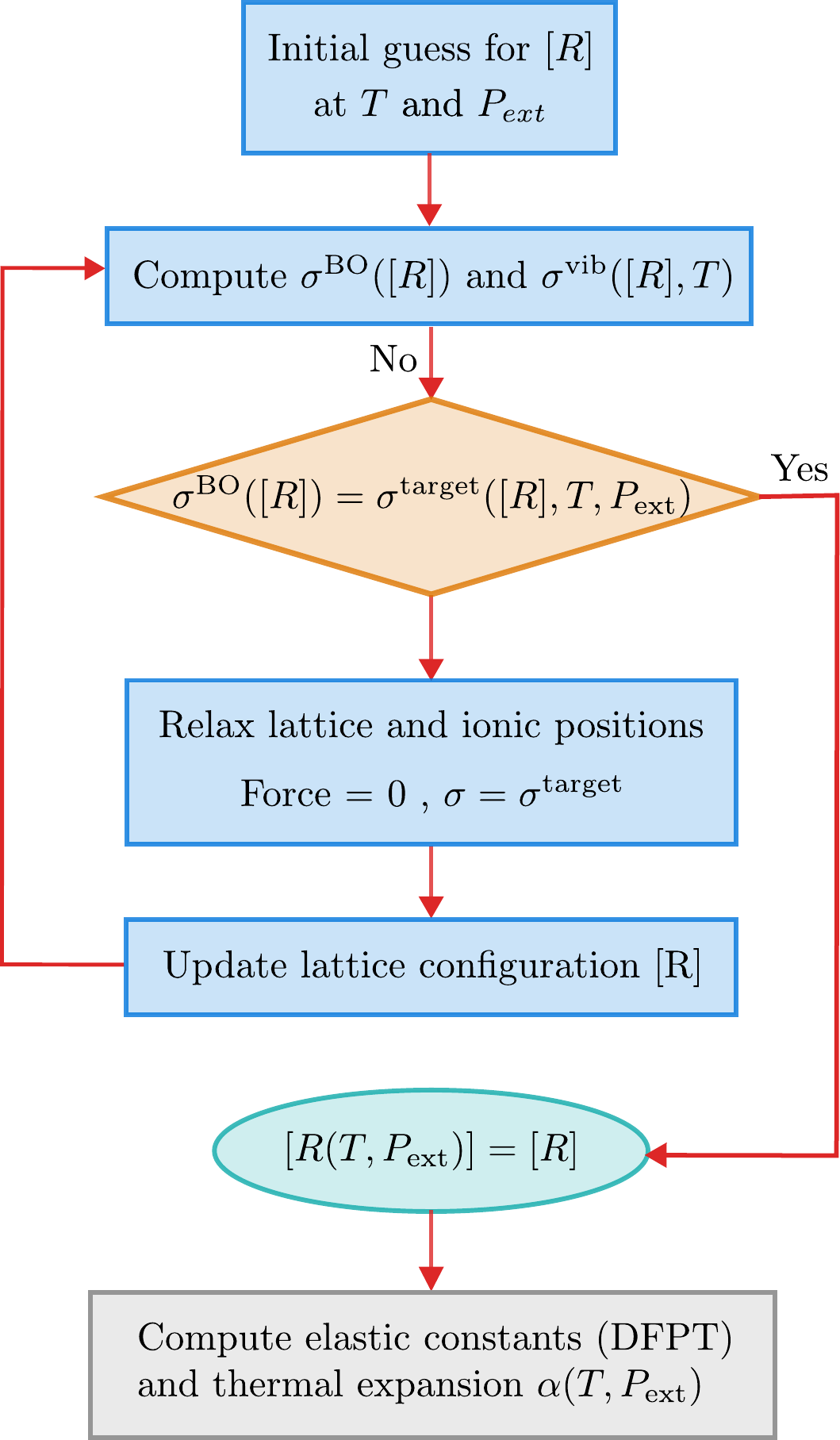}
  \caption{
  Flowchart illustrating the iterative process for determining lattice parameters at temperature $T$ 
  and external pressure $P_{\textrm{ext}}$. The process begins with an initial guess for the lattice 
  configuration $[R]$, followed by the computation of thermal and BO stresses.
  A target stress is defined based on thermal stress and $P_{\textrm{ext}}$, and the lattice and atomic 
  positions are relaxed iteratively until the target stress matches the BO stress, ensuring convergence 
  to the optimal lattice configuration.
   }
		\label{fig:II}
	\end{figure}
After each relaxation, we recompute the thermal stress for the updated lattice and adjust the target 
stress accordingly. 
For each temperature, and (possibly) each external applied pressure, the process is repeated iteratively until the system converges, ensuring that the 
thermal stress and the BO stress match. The calculation flow for finding the lattice parameters \([R(T,P_\textrm{ext})]\) 
is depicted in Fig.~\ref{fig:II}.

\subsection {Thermal expansion}
Once the lattice parameters are obtained at each temperature and pressure, the thermal expansion of different lattice dimensions and angles can be computed using a finite-difference method. 
However, due to minor computational errors introduced by DFT or DFPT calculations, using a standard finite difference method or polynomial interpolation may result in noise in the computed thermal expansion.

To address this, an alternative approach based on Eq.~(\ref{eq:8}) and (\ref{eq:9_}) can be employed to obtain thermal expansion with improved accuracy.
By substituting Eq. (\ref{eq:4}) and Eq. (\ref{eq:22}) into  Eq.~(\ref{eq:8}) with Eq.(\ref{eq:9_}), we can express:
\begin{align}
	\alpha_{\gamma}(T,P_{\textrm{ext}})&=\frac{1}{R_{\gamma}(T,P_{\textrm{ext}})}
    \frac{dR_{\gamma}}{dT}\Bigg|_{[R(T,P_{\textrm{ext}})],T}\\
    =&
    \frac{d\varepsilon_{\gamma}}{dT}\Big|_{[R(T,P_{\textrm{ext}})],T}\\
	=&\sum_{\gamma'} 
	\left[\frac{\partial^2 E_{\textrm{BO}}}{\partial \varepsilon \partial {\varepsilon}'} \Bigg|_{[R(T,P_{\textrm{ext}})]} + \frac{\partial^2 F_{\textrm{vib}}}{\partial \varepsilon \partial {\varepsilon}'} \Bigg|_{[R ^{\bullet}],T}\right .\\
   & \left .+P_{\textrm{ext}}\frac{\partial^2 V}{\partial \varepsilon \partial {\varepsilon}'}\Big|_{[R(T,P_{\textrm{ext}})]} \right]^{-1}_{\gamma \gamma'}\nonumber
	\times \frac{\partial S_{\textrm{vib}}}{\partial \varepsilon_{ \gamma'} } \Big|_{[R(T,P_{\textrm{ext}})],T}.
\end{align}
The first term, which represents the second derivative of the BO energy, can be computed using DFPT calculations of the elasticity tensor~\cite{Hamann2005}.
The elasticity tensor has 6 by 6 elements defined as:
\begin{align}
e^{\textrm{BO}}_{\gamma \gamma'}([R(T,P_{\textrm{ext}})])=&\frac{\partial \sigma^{\textrm{BO}}_{\gamma}([R(T,P_{\textrm{ext}})])}{\partial \varepsilon_{\gamma'}}\\
=&\frac{1}{V(T,P_{\textrm{ext}})}\frac{\partial^2 E_{\textrm{BO}}} {\partial \varepsilon_{\gamma}\partial \varepsilon_{\gamma'}}\Big|_{[R(T,P_{\textrm{ext}})]}. \nonumber
 \end{align}
Consequently, we can express:
 \begin{align}
\frac{\partial^2 E_{\textrm{BO}}} {\partial \varepsilon_{\gamma}\partial \varepsilon_{\gamma'}}\Big|_{[R(T,P_{\textrm{ext}})]}=
V(T,P_{\textrm{ext}}) e^{\textrm{BO}}_{\gamma \gamma'}([R(T,P_{\textrm{ext}})]). 
 \end{align}
where $\gamma$ and $\gamma'$ denote the strain components under consideration, specifically $\{xx, yy, zz, yz, xz, xy\}$.

The second term, representing the second derivative of $F_{\textrm{vib}}$
at the reference configuration 
$[R(T,P_{ext})]$, is computed using finite 
difference methods applied to non-zero strain components. The third term 
corresponds to the second derivative of the volume  with respect to strain 
components at  $ [R(T,P_{ext})]$. For off-diagonal components 
($\gamma\neq\gamma'$, where $\gamma$ and $\gamma'$ are in $\{xx,yy,zz\}$), 
this term equals $V(T,P_{ext})$, while all other terms are zero.
$S_{\textrm{vib}}$ is determined from Eq.(\ref{eq:9_}) on the same set of points as for the vibrational free energy. This allows its quadratic interpolation as a function of the strain tensor, and hence the computation of its derivative with respect to the strain tensor.

Thus, after determining $[R[T, P_{\textrm{ext}}]]$, a single DFPT calculation at the $\Gamma$ point is sufficient to obtain the thermal expansion at each $T$ and $P_{\textrm{ext}}$.

\subsection { Elastic constants}
In addition to thermal expansion, the elastic constants at different temperatures and pressures can also be derived. The elastic constants are defined as:
\begin{align}
e_{\gamma \gamma'}([R(T,P_{\textrm{ext}})],T,P_{\textrm{ext}})
&=
\nonumber\\
\frac{1}{V(T,P_{\textrm{ext}})}&
\frac{\partial^2 G} {\partial \varepsilon_{\gamma}\partial \varepsilon_{\gamma'}}\Big|_{[R(T,P_{\textrm{ext}})],T,P_{\textrm{ext}}}. 
 \end{align}
We can express the free energy as a sum of the vibrational and BO energies:
 
\begin{align}
\frac{\partial^2 G}{\partial \varepsilon_{\gamma}\partial \varepsilon_{\gamma'}} &\Bigg|_{[R(T,P_{\textrm{ext}})],T,P_{\textrm{ext}}}=
\frac{\partial^2 E_{\textrm{BO}}}{\partial \varepsilon_{\gamma}\partial \varepsilon_{\gamma'}} \Bigg|_{[R(T,P_{\textrm{ext}})]} \\
&+ \frac{\partial^2 F_{\textrm{vib}}}{\partial \varepsilon_{\gamma}\partial \varepsilon_{\gamma'}} \Bigg|_{[R ^{\bullet}],T}+P_{\textrm{ext}}\frac{\partial^2 V}{\partial \varepsilon_{\gamma} \partial {\varepsilon_{\gamma'}}}\Big|_{[R(T,P_{\textrm{ext}})]}. \nonumber
\end{align}

Considering the symmetry constraints discussed in earlier sections, second derivatives of the vibrational free energy are omitted in cases where the first derivatives vanish. However, to determine the complete set of thermal elastic constants, additional calculations are necessary to account for all non-zero elements of the elastic tensor. To achieve this, additional deformations and phonon spectra calculations must be performed.
Moreover, the deformations required for thermal expansion may differ from 
those needed to determine elastic constants.
The distinction lies in the fact that, for thermal expansion, the use of crystal symmetries allows for a significant reduction in the number of required phonon spectra calculations.

For cubic and uniaxial structures, thermal expansion can be treated as 
a one- or two-dimensional problem, requiring only 3 and 6 deformations, 
respectively. In contrast, determining elastic constants necessitates treating 3 DOFs, similar to orthorhombic crystals, to fully 
evaluate the elastic tensor for {$xx$, $yy$, $zz$} components. Despite this, the number 
of calculations remains lower than that for orthorhombic structures due to 
symmetry considerations. Specifically, the 10 deformations needed for 
orthorhombic systems reduce to 4 and 7 for cubic and uniaxial crystals, 
respectively.

If the objective is to determine thermal expansion alongside elastic constants, 
the deformations required for elastic constants should be employed from the
beginning. Table~\ref{tab:II} summarizes the non-zero elastic constants for 
various crystal systems and the number of deformations required to compute them.

\begin{table}
\caption{List of non-zero elastic constants and the corresponding number of required deformations for various crystal systems.}
\label{tab:II}

	\begin{ruledtabular}
		\begin{tabular}{lccc}
   Crystal & Elastic constant &  \#deform\\
  \hline
  Cubic & $C_{11}, \, C_{12}, \, C_{44}$  & 6 \\
  \hline
  Hexagonal & $C_{11}, \, C_{12}, \, C_{13}, \, C_{33}, \, C_{44}$ &9\\
   \hline
  \begin{tabular}{l} 
   Trigonal \\(Rhombohedral) \end{tabular}& $C_{11}, \, C_{12}, \, C_{13}, \, C_{14}, \, C_{33}, \, C_{44}$&10\\ 
     \hline
 Tetragonal & $ C_{11}, \, C_{12}, \, C_{13}, \, C_{33}, \, C_{44}, \, C_{66}$&11\\
  \hline
Orthorhombic & \begin{tabular}{c} $C_{11}, \, C_{12}, \, C_{13}, \, C_{22}, \, C_{23}, $ \\ $ C_{33}, \, C_{44}, \, C_{55}, \, C_{66} 
   $\end{tabular} &16 \\
			\hline
			 Monoclinic  &  \begin{tabular}{c} $ 
              C_{11}, \, C_{12}, \, C_{13}, \, C_{15}, \, C_{22}, \, C_{23}, \, C_{25}, $\\$ C_{33}, \, C_{35}, \, C_{44}, \, C_{46}, \, C_{55}, \, C_{66} $ 
       \end{tabular} & 18\\
    \hline
			Triclinic  &  \begin{tabular}{l} $ C_{ij} \quad \text{for } i, j = 1, 2, \dots, 6 $ \end{tabular}  & 28
     \end{tabular}
	\end{ruledtabular}
\end{table}

Here, we present the equations for two cases, orthorhombic 
and monoclinic. The equations for the remaining
crystal and slab structures, as listed in Table~\ref{tab:I}, are provided in the Appendix.

\subsection{Orthorhombic case}
For sake of compactness, in this subsection and the following one, the $P_{\textrm{ext}}$ dependence of the components of the lattice vectors is not explicitly indicated, unlike their temperature dependence. For the orthorhombic case, given the strain matrix $[\varepsilon^{\textrm{BO}}]$, with only diagonal elements, the lattice vectors can be determined as follows:
  \begin{align}
  \label{eq:R_orthorhombic}
		[R]&=	 \left[ \begin{array}{lll}
		(1+\varepsilon^{\textrm{BO}}_{xx}) R^{\textrm{BO}}_{1,x}
		&(1+\varepsilon^{\textrm{BO}}_{xx}) R^{\textrm{BO}}_{2,x}
		&  (1+\varepsilon^{\textrm{BO}}_{xx})  R^{\textrm{BO}}_{3,x}  \\
		(1+\varepsilon^{\textrm{BO}}_{yy})  R^{\textrm{BO}}_{1,y}
		&  (1+\varepsilon^{\textrm{BO}}_{yy}) R^{\textrm{BO}}_{2,y}
		&  (1+\varepsilon^{\textrm{BO}}_{yy})  R^{\textrm{BO}}_{3,y}  \\
		(1+\varepsilon^{\textrm{BO}}_{zz}) R^{\textrm{BO}}_{1,z}
		& (1+\varepsilon^{\textrm{BO}}_{zz}) R^{\textrm{BO}}_{2,z}
		& (1+\varepsilon^{\textrm{BO}}_{zz})  R^{\textrm{BO}}_{3,z}
	\end{array} \right].
 \end{align} 
 Note that this accounts for different orthorhombic Bravais lattices (also centered/face centered ones), for which the primitive vectors might not be aligned with the $x$, $y$ or $z$ directions.
 The use of the primitive cell is the most economical for DFT and DFPT calculations, but
 it is easier to focus on conventional directions 
 $x$, $y$ or $z$ to characterize the temperature dependence of the geometry of the crystal. For this purpose, we define new parameters $A_x$, $B_y$, and
 $C_z$ that represent the lengths of the components of the lattice vectors in the $x$, $y$, and
 $z$ directions. These are calculated as:
 \begin{align}
 A_x= |R_{1,x}|+|R_{2,x}|+|R_{3,x}|,\nonumber\\
  B_y= |R_{1,y}|+|R_{2,y}|+|R_{3,y}|,\\
   C_z= |R_{1,z}|+|R_{2,z}|+|R_{3,z}|.\nonumber
	\end{align} 
The strain components relative to the unstrained (Born-Oppenheimer) configuration can
subsequently be determined using the following expressions:
\begin{align}
		\varepsilon^{\textrm{BO}}_{xx}(T)= \frac{A_x(T)}{A^{\textrm{BO}}_x}-1 \quad,\quad \varepsilon^{\textrm{BO}\bullet}_{xx}= \frac{A^{\bullet}_x}{A^{\textrm{BO}}_x}-1,\nonumber\\
		\varepsilon^{\textrm{BO}}_{yy}(T)= \frac{B_y(T)}{B^{\textrm{BO}}_y}-1 \quad,\quad \varepsilon^{\textrm{BO}\bullet}_{yy}= \frac{B^{\bullet}_y}{B^{\textrm{BO}}_y}-1,\\
		\varepsilon^{\textrm{BO}}_{zz}(T)= \frac{C_z(T)}{C^{\textrm{BO}}_z}-1 \quad,\quad \varepsilon^{\textrm{BO}\bullet}_{zz}= \frac{C^{\bullet}_z}{C^{\textrm{BO}}_z}-1.\nonumber
	\end{align}
Employing Eq.~(\ref{eq:30}) the thermal stress can be computed as:
\begin{align}
\label{eq:Stress_orthorhombic}
 \sigma^{\textrm{vib}}_{xx}\Big|_{[R(T)]}
 =&   \frac{1}{V(T)}\frac{dF_{\textrm{vib}}}{d\varepsilon^{\textrm{BO}}_{xx}}\Big|_{[R(T)],T}
 \left(1+\varepsilon^{\textrm{BO}}_{xx}(T)\right)\nonumber\\
		\sigma^{\textrm{vib}}_{yy}\Big|_{[R(T)]}
        =&   \frac{1}{V(T)}\frac{dF_{\textrm{vib}}}{d\varepsilon^{\textrm{BO}}_{yy}}\Big|_{[R(T)],T}
 \left(1+\varepsilon^{\textrm{BO}}_{yy}(T)\right)\\
		\sigma^{\textrm{vib}}_{zz}\Big|_{[R(T)]}
        =&   \frac{1}{V(T)}\frac{dF_{\textrm{vib}}}{d\varepsilon^{\textrm{BO}}_{zz}}\Big|_{[R(T)],T}
 \left(1+\varepsilon^{\textrm{BO}}_{zz}(T)\right)\nonumber
  \end{align}

In these equations, scaling factors (${A_x(T)}/{A^{\textrm{BO}}_x}$, ${B_y(T)}/{B^{\textrm{BO}}_y}$, and 
${C_z(T)}/{C^{\textrm{BO}}_z}$) are applied because the strains are calculated relative to the equilibrium 
lattice $[R^{\textrm{BO}}]$, while the thermal stresses require derivatives with respect to the lattice at $[R(T)]$, as described previously.
For orthorhombic structures, the equations provided are general and applicable to any choice of primitive cell.
\subsection{Monoclinic case}
For the monoclinic case, to simplify the calculations and corresponding implementation, we use a standardized definition of the primitive cell.
In this standardized definition, the lattice has four degrees of freedom: $\varepsilon^{\textrm{BO}}_{xx}$, $\varepsilon^{\textrm{BO}}_{yy}$,
$\varepsilon^{\textrm{BO}}_{zz}$, and $\varepsilon^{\textrm{BO}}_{xz}$. 
Correspondingly, the lattice vectors for a simple monoclinic primitive cell can be defined as follows: 
\begin{align} \label{eq:R-monoclinic}
		&[R]= 
	\left[ \begin{array}{lll}
	R_{1,x} & 	0 & 	R_{3,x}  \\ 
		0 & 	R_{2,y} & 	0 \\
			0 & 	0 & 	R_{3,z}  
\end{array} \right]\\
&~~~~=
\left[ \begin{array}{lll}
	1+\varepsilon^{\textrm{BO}}_{xx} & 	0 & 	\varepsilon^{\textrm{BO}}_{xz} \\ 
		0 & 	1+\varepsilon^{\textrm{BO}}_{yy} & 	0 \\
	0& 	0 & 	1+\varepsilon^{\textrm{BO}}_{zz}
\end{array} \right].[R^{\textrm{BO}}]=\nonumber\\ 
&\left[ \begin{array}{lll}
		(1+\varepsilon^{\textrm{BO}}_{xx}) R^{\textrm{BO}}_{1,x}
		&  0 
		&(1+\varepsilon^{\textrm{BO}}_{xx}) R^{\textrm{BO}}_{3,x}+  \varepsilon^{\textrm{BO}}_{xz}    R^{\textrm{BO}}_{3,z}  \\
		0
		&  (1+\varepsilon^{\textrm{BO}}_{yy}) R^{\textrm{BO}}_{2,y}
		& 0  \\
		0
		&  0
		&   (1+\varepsilon^{\textrm{BO}}_{zz})  R^{\textrm{BO}}_{3,z} 
	\end{array} \right]\nonumber
	\end{align} 
Therefore, the strain components for a monoclinic structure can be defined as:
\begin{align}
		\varepsilon^{\textrm{BO}}_{xx}(T)=& \frac{R_{1,x}(T)}{R_{1,x}^{\textrm{BO}}}-1 \quad,\quad \varepsilon^{\textrm{BO}\bullet}_{xx}= \frac{R_{1,x}^{\bullet}}{R_{1,x}^{\textrm{BO}}}-1\nonumber\\
		\varepsilon^{\textrm{BO}}_{yy}(T)=& \frac{R_{2,y}(T)}{R_{2,y}^{\textrm{BO}}}-1 \quad,\quad \varepsilon^{\textrm{BO}\bullet}_{yy}= \frac{R_{2,y}^{\bullet}}{R_{2,y}^{\textrm{BO}}}-1\nonumber\\
		\varepsilon^{\textrm{BO}}_{zz}(T)=& \frac{R_{3,z}(T)}{R_{3,z}^{\textrm{BO}}}-1 \quad,\quad \varepsilon^{\textrm{BO}\bullet}_{zz}= \frac{R_{3,z}^{\bullet}}{R_{3,z}^{\textrm{BO}}}-1\nonumber\\
  \varepsilon^{\textrm{BO}}_{xz}(T)=& \frac{R_{3,x}(T) R_{1,x}^{\textrm{BO}}-R_{1,x}(T) R_{3,x}^{\textrm{BO}}}{R_{1,x}^{\textrm{BO}} R_{3,z}^{\textrm{BO}}}\\
  \varepsilon^{\textrm{BO}\bullet}_{xz}=&  \frac{R_{3,x}^{\bullet} R_{1,x}^{\textrm{BO}}-R_{1,x}^{\bullet} R_{3,x}^{\textrm{BO}}}{R_{1,x}^{\textrm{BO}} R_{3,z}^{\textrm{BO}}}\nonumber
	\end{align}
and we can obtain the thermal stress as follow:
\begin{align}
& \sigma^{\textrm{vib}}_{xx}\Big|_{[R(T)]}=  \frac{1}{V(T)}\frac{dF_{\textrm{vib}}}{d\varepsilon^{\textrm{BO}}_{xx}}\Big|_{[R(T)],T}\left(1+\varepsilon^{\textrm{BO}}_{xx}(T)\right)\nonumber\\
	&	\sigma^{\textrm{vib}}_{yy}\Big|_{[R(T)]}=  \frac{1}{V(T)}\frac{dF_{\textrm{vib}}}{d\varepsilon^{\textrm{BO}}_{yy}}\Big|_{[R(T)],T}\left(1+\varepsilon^{\textrm{BO}}_{yy}(T)\right)\nonumber\\
	&	\sigma^{\textrm{vib}}_{zz}\Big|_{[R(T)]}=   \frac{1}{V(T)}\frac{dF_{\textrm{vib}}}{d\varepsilon^{\textrm{BO}}_{zz}}\Big|_{[R(T)],T}\left(1+\varepsilon^{\textrm{BO}}_{zz}(T)\right)\nonumber\\
& \sigma^{\textrm{vib}}_{xz}\Big|_{[R(T)]}=  \frac{1}{V(T)}
  \left(\frac{dF_{\textrm{vib}}}{d\varepsilon^{\textrm{BO}}_{xz}}\Big|_{[R(T)],T}\left(1+\varepsilon^{\textrm{BO}}_{zz}(T)\right)\right .\nonumber
 \\
  &+\left . \frac{dF_{\textrm{vib}}}{d\varepsilon^{\textrm{BO}}_{xx}}\Big|_{[R(T)],T} 
 \varepsilon^{\textrm{BO}}_{xz}(T) \right).
 \label{eq:stress_monoclinic}
  \end{align}

\section{Computational details}\label{sec:Computational}

The calculations for ground-state energies were performed using Density Functional
Theory (DFT), while phonon frequencies were determined via Density-Functional 
Perturbation Theory (DFPT). Spin-orbit interactions were not included in these 
simulations. Optimized norm-conserving Vanderbilt pseudopotentials~\cite{Hamann2013} 
carefully validated against all-electron full-potential methods~\cite{Lejaeghere2016, Bosoni2023}
were sourced from the 
\verb|Pseudo-Dojo| project~\cite
{VanSetten2018}, and the exchange-correlation
effects were described using the GGA-PBEsol functional~\cite{Perdew2008}.  
Lattice parameters and atomic positions were optimized iteratively until forces on 
atoms were less than \(10^{-5}\) Hartree/Bohr\(^3\) and stress components were 
below \(10^{-8}\) Hartree/Bohr\(^3\).  
To produce smooth energy-volume curves, an energy cutoff smearing parameter of 1.0 
Ha was applied~\cite{Laflamme2016}.  
Brillouin zone integrations were carried out with carefully chosen wavevector 
grids 
ensuring that errors in total energy remained under 1 
meV per atom. The specific parameters used for each material are listed in 
Table~\ref{tab:III}.  
All computations were executed using the ABINIT software suite (version 
9.10.3)~\cite{Gonze2002,Gonze2020,Romero2020}. The phonon density of states 
(PHDOS) was calculated through a Gaussian broadening approach
with a smearing value of 1 cm$^{-1}$ (approximately 
$4.5\times10^{-6}$ Hartree),   
which is the default value for ABINIT versions above v9.10.

The reference structure, $[R^{\bullet}]$ where a uniform strain shift is applied to the diagonal components. Specifically, the strain components are set as $\varepsilon^{\textrm{BO}\bullet}_{xx} = \varepsilon^{\textrm{BO}\bullet}_{yy} = \varepsilon^{\textrm{BO}\bullet}_{zz} = 0.005$. 
However, no shift is applied to the off-diagonal strain components ($\varepsilon^{\textrm{BO}\bullet}_{xy}, \varepsilon^{\textrm{BO}\bullet}_{xz}, \varepsilon^{\textrm{BO}\bullet}_{yz}$) since their thermal variation is not well understood. These components are related to changes in lattice angles rather than direct expansion or contraction, making it uncertain how they should be adjusted for thermal expansion.

\subsection{Materials}

We analyze the thermal properties of 12 materials using a combination of the
ZSISA method and the approximation approach ${\textrm{E}{\infty} \textrm{Vib}2}$
to account for anisotropic behavior. 
These materials encompass a wide range of crystallographic symmetries, providing
a robust framework for investigating anisotropic thermal properties. 
Among the cubic systems, MgO was studied, while ZnO, GaN, and AlN represented
the hexagonal group. Trigonal materials included \ce{CaCO3} 
and \ce{Al2O3}, and tetragonal systems were exemplified by PbO and \ce{SnO2}.
In the orthorhombic group, \ce{YAlO3} was analyzed, whereas \ce{ZrO2}, \ce{HfO2}, 
and \ce{MgP4} belong to the monoclinic category. 
Finally, the triclinic group was represented by \ce{Al2SiO5}. This diverse
selection spans the principal crystallographic space groups, enabling a
comprehensive study of the anisotropic thermal properties across various regions. 
material classes. 

For cubic systems, simpler approaches such as the v-ZSISA-QHA method or similar
approximations~\cite{Rostami2024}
may be preferred to calculate the thermal expansion,
as the new method does not offer significant advantages in this regard. However,
when it comes to computing elastic constants, the utility of this new method becomes apparent. 
Therefore, for MgO, we focused on calculating the elastic constants to demonstrate its applicability and effectiveness.

We did not generate QHA results for all materials due to their significant 
computational expense and resource demands. As an initial test of the proposed 
approximations, we applied the full ZSISA-QHA method to a few uniaxial systems, 
with ZnO presented here as a representative case.

For materials with more than two lattice degrees of freedom, the computational cost of 
phonon spectra calculations increases considerably, even taking into account (reduced) symmetries. Consequently, testing the ZSISA-QHA 
method for those systems with lower symmetries was not undertaken. 
For these cases, we nevertheless performed an internal check 
inside the  ${\textrm{E}{\infty} \textrm{Vib}2}$ method, comparing the thermal stress approach with the results obtained from fitting the
high-dimensional free energy.
For monoclinic \ce{ZrO2}, which has four degrees of freedom, we constructed a 4D surface of 
free energies. This required 625 BO energy 
evaluations for  $\textrm{E}{\infty}$ while the second degree Taylor expansion for the phonon free energy relied on 15 phonon spectra calculations for  
${F_{\textrm{Vib}2}}$. The results were consistent between both approach, with a demonstrated high accuracy. These results emphasize the practicality and 
reliability of the approach for systems with complex anisotropic properties.

\small{
\begin{table*}
\caption{For each material, the table presents the space group (Hermann–Mauguin notation), lattice parameters obtained from DFT calculations or measured experimentally at room temperature, angles between primitive cell vectors (``$\angle$") as determined by DFT and experiments, the number of lattice degrees of freedom (lDOF), the number of atoms in the primitive cell, and the number of internal degrees of freedom (iDOF). Additionally, computational parameters such as the plane-wave energy cutoff ($E_{\textrm{cut}}$), electronic wavevector sampling (k-grid), and vibrational wavevector sampling (q-grid) are provided. \label{tab:III}}
\begin{ruledtabular}
    \begin{tabular}{c|cccccccccc}
 Material &  Group &\multicolumn{2}{c}{Lattice  (\AA) }  &\multicolumn{2}{c}{$\angle$($^\circ$)}  & lDOF &{\#}atoms & iDOF & $E_\textrm{cut}$(Ha) & k-grid/ \\
 \cline{3-4} \cline{5-6}
   &     &DFT  &Exp.  &DFT  &Exp.  & &&&& q-grid \\
       
        \hline
        Cubic:\\
        \ce{MgO  }&$Fm\bar{3}m $&  a= 4.214  & a=4.212~\cite{Li2006} &$\alpha$= 90      &$\alpha$= 90              &1 &2&0&60& 8 $\times$8 $\times$8  \\

 \hline
 Hexagonal :\\
        \ce{ZnO  }&$P6_3mc$&\begin{tabular}{c}  a= 3.227\\ c= 5.206  \\~ \end{tabular}  &\begin{tabular}{l} a=
3.250~\cite{Karzel1996} \\ c= 5.204 \\~ \end{tabular}            &\begin{tabular}{c} $\alpha$= 90 \\  $\gamma$= 120  \\~ \end{tabular}   &\begin{tabular}{c} $\alpha$= 90 \\  $\gamma$= 120  \\~ \end{tabular}          &2&4&1&42& 6 $\times$6 $\times$4  \\

        \ce{AlN  }&$P6_3mc$&\begin{tabular}{l}  a= 3.113 \\ c= 4.982  \\~ \end{tabular}   &\begin{tabular}{l} a= 3.110~\cite{SCHULZ1977}\\c=4.980\\~\end{tabular}     
&\begin{tabular}{c} $\alpha$= 90 \\  $\gamma$= 120  \\~ \end{tabular}  &\begin{tabular}{c} $\alpha$= 90 \\  $\gamma$= 120  \\~ \end{tabular}
&2 &4&1&40& 6 $\times$6 $\times$4  \\

\ce{GaN  }&$P6_3mc$&\begin{tabular}{c}  a= 3.184\\ c= 5.186  \\~ \end{tabular}  &\begin{tabular}{l} a= 3.190~\cite{SCHULZ1977}\\c=5.189\\~\end{tabular}  &\begin{tabular}{c} $\alpha$= 90 \\  $\gamma$= 120  \\~ \end{tabular}
  &\begin{tabular}{c} $\alpha$= 90 \\  $\gamma$= 120  \\~ \end{tabular}
&2 &4&1&40& 6 $\times$6 $\times$4  \\

  \hline
  Trigonal:\\
\ce{CaCO3}&$R\bar{3}c  $& a= 6.313 & a=6.344~\cite{wang2018} &$\alpha$= 46.58 &$\alpha$~\cite{wang2018}= 46.31        &2  &10&1&42& 5 $\times$5 $\times$5   \\~\\

\ce{Al2O3}&$R\bar{3}c  $&  a= 5.133 &
 a=5.129~\cite{GRABOWSKI2018}        &$\alpha$= 55.35 &$\alpha$~\cite{GRABOWSKI2018}=55.28     &2  &10&2&42& 4 $\times$4 $\times$4   \\
    \hline
 Tetragonal: \\
  \ce{SnO2}& $P4_2/mnm$& \begin{tabular}{c}  a= 4.777 \\ c= 3.221  \end{tabular} & \begin{tabular}{c}  a= 4.737~\cite{Haines1997} \\ c= 3.186   \end{tabular}  &$\alpha$= 90   &$\alpha$= 90             &2 &6&1&45& 5 $\times$5 $\times$7  \\\\
  \hline
  Orthorhombic:\\
\ce{YAlO3}&$Pnma  $&\begin{tabular}{c}  a= 5.322\\ b= 7.358 \\ c= 5.158  \\~ \end{tabular}&\begin{tabular}{l}  a=
5.329~\cite{Aggarwal2005}\\ b= 7.371\\ c= 5.180  \\~ \end{tabular}&$\alpha$= 90   &$\alpha$= 90             &3  &20&7&45& 3 $\times$2 $\times$3  \\
	\hline
			 Monoclinic: \\
\ce{ZrO2 }&$P2_1/c$&\begin{tabular}{c}  a= 5.126\\ b= 5.209 \\ c= 5.294  \\~ \end{tabular}&\begin{tabular}{l}   a=
5.169~\cite{McCullough1959}\\ b= 5.232 \\ c= 5.341  \\~ \end{tabular}&\begin{tabular}{c}$\alpha$= 90 \\ $\beta$= 99.59\\~\end{tabular}  &\begin{tabular}{c}$\alpha$= 90 \\ $\beta$=  99.25\\~ \end{tabular} &4&12&9&42& 4 $\times$4 $\times$4 \\

\ce{HfO2 }&$P2_1/c$&\begin{tabular}{c}  a= 5.077 \\ b= 5.155 \\ c= 5.252  \\~ \end{tabular}&\begin{tabular}{l}   a= 5.116~\cite{Haggerty2014}
\\ b= 5.179 \\ c= 5.289  \\~ \end{tabular}&\begin{tabular}{c}$\alpha$= 90 \\ $\beta$= 99.64 \\~ \end{tabular} &\begin{tabular}{c}$\alpha$= 90 \\ $\beta$= 99.25 \\~ \end{tabular} & 4&12&9&42& 4 $\times$4 $\times$4    \\     

        \ce{MgP4}&$P2_1/c$ &\begin{tabular}{c}  a= 5.132 \\ b= 5.055 \\ c= 7.529  \\~ \end{tabular}&\begin{tabular}{l}   a= 5.15
~\cite{ElMaslout1975}\\ b= 5.10 \\ c= 7.50   \\~ \end{tabular}&\begin{tabular}{c}$\alpha$= 90 \\ $\beta$= 98.54\\~\end{tabular} &\begin{tabular}{c}$\alpha$= 90 \\$\beta$= 81 \\~ \end{tabular} &4&10&6&42& 4 $\times$4 $\times$3  \\ 
    \hline
			Triclinic:  \\
            \ce{Al2SiO5}&$P\overline{1}$&\begin{tabular}{c}  a= 5.577 \\ b= 7.130 \\ c= 7.864  \\~ \end{tabular}&\begin{tabular}{l}   a= 5.569~\cite{Fortes2019}
\\ b=7.116\\ c= 7.844   \\~ \end{tabular}
&\begin{tabular}{c}$\alpha$= 73.99  \\ $\beta$=  90.02  \\$\gamma$= 78.89 \\~ \end{tabular}
&\begin{tabular}{cc}$\alpha$=74.00\\  $\beta$=  89.99 \\
 $\gamma$= 78.88 \\~ \end{tabular}

&6&32&48&42& 3 $\times$2 $\times$2  
        \label{tab:tab2}
    \\~ \end{tabular}
\end{ruledtabular}
\end{table*}}

   \section{Results and Discussions}\label{sec:result}
   The proposed method was tested on various materials representing a wide range of 
   crystallographic structures, from cubic to triclinic.
For cubic structures, applying this thermal expansion method using 
thermal stress does not offer significant advantages over the
v-ZSISA-${\textrm{E}{\infty} \textrm{Vib}2}$ approach introduced in our previous work~\cite{Rostami2024}. 
In fact, the computational effort may exceed that of the previous 
method. However, the current approach proves beneficial for 
computing elastic constants at finite temperatures and pressures, 
requiring only five phonon spectra calculations to determine the three independent elastic constants (two more than for the volume only). To illustrate this, we used \ce{MgO} as a representative 
cubic material for elastic constant calculations.

For uniaxial systems, including hexagonal, trigonal, and tetragonal 
structures, we computed thermal expansion for several compounds. 
These include ZnO, GaN, and AlN in the wurtzite structure (space 
group $P6_3mc$), \ce{CaCO3} and \ce{Al2O3} in the rhombohedral 
structure (space group $R\bar{3}c$), and \ce{SnO2} in the tetragonal
structure (space group $P4_2/mnm$), as summarized in Table~\ref{tab:III}.

Orthorhombic \ce{YAlO3} was selected as a representative material 
for orthorhombic structures. In the monoclinic category, we examined 
\ce{ZrO2}, \ce{HfO2}, and \ce{MgP4}, all with space group $P2_1/c$. 
Finally, \ce{Al2SiO5} (space group $P\bar{1}$) was chosen as a 
representative of triclinic symmetry. The number of atoms in the 
primitive unit cell for each structure is also provided in Table~\ref{tab:III}.
 The reference structure includes base strains of 
$\varepsilon^{\textrm{BO}}_{xx} = \varepsilon^{\textrm{BO}}_{yy} = \varepsilon^{\textrm{BO}}_{zz} = 0.005$ 
and $\varepsilon^{\textrm{BO}}_{xz} =\varepsilon^{\textrm{BO}}_{yz} =\varepsilon^{\textrm{BO}}_{xy}= 0$,
consistent with the positive thermal expansion behavior of our materials. 
 In our calculations, $[R^{\textrm{BO}}]$ is chosen as the relaxed BO configuration at zero strain 
 and zero pressure, serving as the reference point for all deformation calculations.
 We used a total of 20 temperature points to obtain our results, employing an adaptive step size
 strategy to balance precision and efficiency. From 0 K to 200 K, where thermal behavior changes
 more rapidly, a finer step size of 25 K was chosen. Between 200 K and 500 K, a step size of 50 K 
 was used, while for temperatures above 500 K up to 1000 K, a larger step size of 100 K was 
 sufficient due to the smoother thermal response at higher temperatures. 

In this work, we present detailed results for MgO, ZnO, and 
\ce{ZrO2}. The results for the remaining materials are included in the Supplemental Material (SM)~\cite{SM}.
\subsection{MgO}
MgO, with its cubic structure, has a lattice constant of a=4.214 \AA ~ in our computational 
setup. For thermal expansion calculations, only one degree of freedom governs the 
lattice, making the process equivalent to the 
${\textrm{v-ZSISA-E}{\infty} \textrm{Vib}2}$ method from our previous work.
This approach simplifies the procedure by interpolating the total energy at different 
volumes with selected $E_{\textrm{BO}}$ values to fit an equation of state (EOS), 
eliminating the need for a complete workflow to determine $E_{\textrm{BO}}$.
In contrast, computing elastic constants requires treating the cubic primitive cell as monoclinic to capture
all necessary distortions and determine each non-zero element of the elastic constants tensor.
There are important differences when applying monoclinic deformations to cubic systems.
Due to symmetry, the deformations along the $\varepsilon_{xx}$,  $\varepsilon_{yy}$, and 
$\varepsilon_{zz}$ directions are equivalent, as are the shear strains  
 $\varepsilon_{xy}$,  $\varepsilon_{xz}$, and  $\varepsilon_{yz}$.
  This symmetry reduces the number of required deformations. Furthermore, since the first
  derivative of the free energy, and consequently the stress, is zero for non-orthogonal
  directions ($xz$, $xy$, $yz$), it is possible to use orthorhombic equations for computing $[R(T,P_\textrm{ext})]$,
  simplifying the overall process. 
  
  However, to compute $C_{44}$, one additional equation is necessary to find the second derivative 
  ${\partial^2 F_{\textrm{vib}}}/{\partial \varepsilon^{\textrm{BO2}}_{xz} }$. 
  Unlike the monoclinic case, where symmetric deformations $\pm \delta \varepsilon^{\textrm{BO}}_{xz}$
  are used, cubic symmetry necessitates applying both $\delta \varepsilon^{\textrm{BO}}_{xz}$ and
  $2 \delta \varepsilon^{\textrm{BO}}
  _{xz}$. This distinction arises from the equivalence of symmetric strain variations 
  due to the zero derivative of $F_{\textrm{vib}}$ in the $xz$ direction and is further explained in the
  appendix, where the unique deformation requirements for cubic and uniaxial 
  systems are discussed.
The applied strains for elastic constant calculations were 
$[\varepsilon^{\textrm{BO}\bullet}]$,
$([\varepsilon^{\textrm{BO}\bullet}] \pm \delta \varepsilon^{\textrm{BO}}_{xx})$,
$([\varepsilon^{\textrm{BO}\bullet}] -\delta \varepsilon^{\textrm{BO}}_{xx} -\delta \varepsilon^{\textrm{BO}}_{yy})$,
$([\varepsilon^{\textrm{BO}\bullet}] + \delta \varepsilon^{\textrm{BO}}_{xz})$, and
$([\varepsilon^{\textrm{BO}\bullet}] + 2\delta \varepsilon^{\textrm{BO}}_{xz})$ with 
$\delta \varepsilon^{\textrm{BO}}=0.005$.

\begin{figure}[!]
\includegraphics[width=0.45\textwidth]{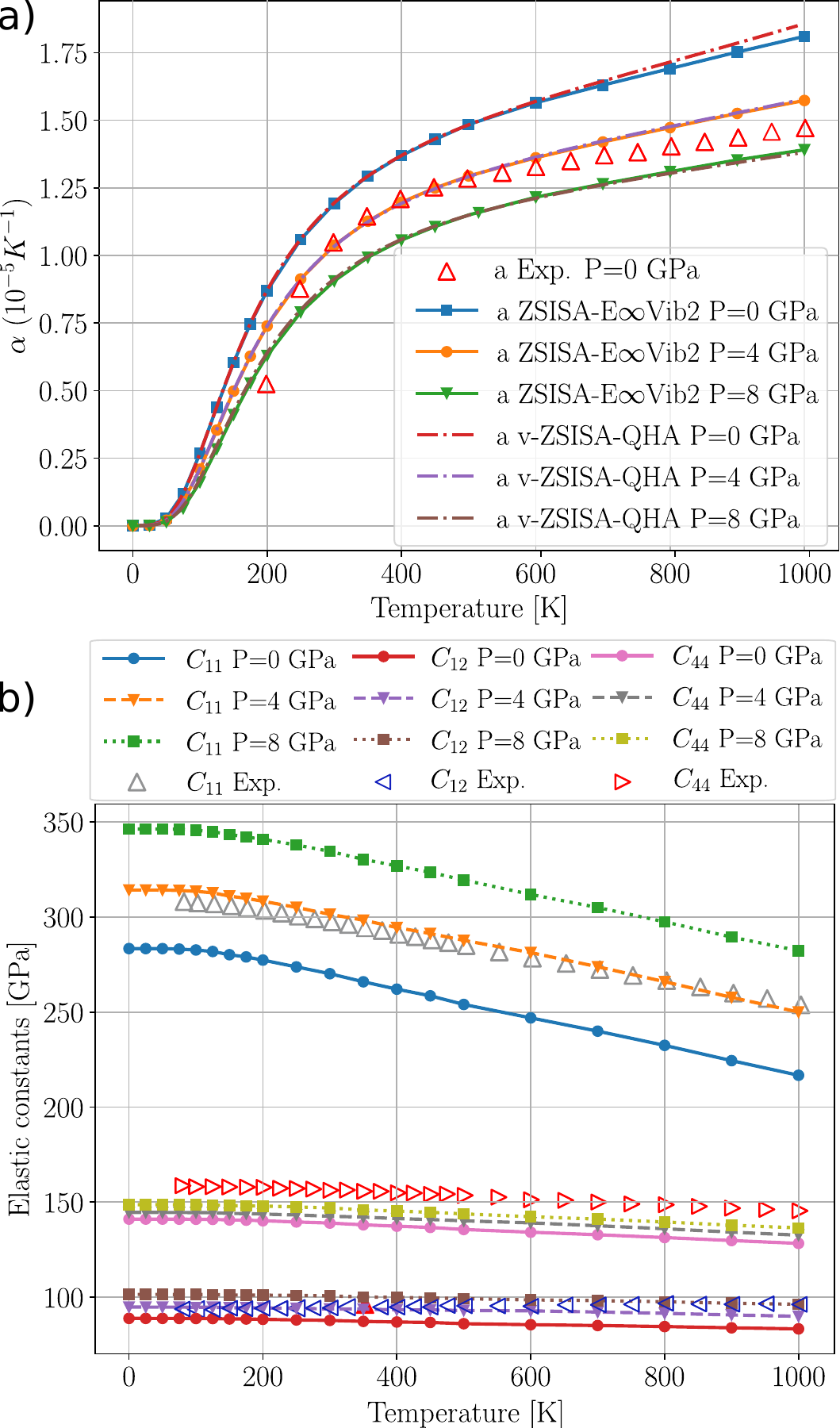}
  \caption{Thermal expansion and elastic constants of MgO as a function of temperature. 
  (a) Thermal expansion at external pressures $P_{\textrm{ext}}$=0, 4, and 8 GPa, 
  computed using v-ZSISA-QHA (dashed lines) and the proposed orthorhombic approach 
  (solid lines with markers). Experimental data~\cite{Dubrovinsky1997} at $P_{\textrm{ext}}=0$ GPa are shown as discrete
  points for comparison. (b) Temperature dependence of the elastic constants $C_{11}$, $C_{12}$, 
  and $C_{44}$ at $P_{\textrm{ext}}$=0 GPa (solid lines), 4 GPa (dashed lines), and 8 GPa
  (dotted lines with markers). Experimental values~\cite{Sumino1983} at $P_\textrm{ext}=0$ GPa are represented by discrete points.}
  \label{fig:MgO}
\end{figure}

Figure~\ref{fig:MgO}  presents the thermal expansion and elastic constants of MgO 
as a function of temperature at different external pressures.
The top panel compares the thermal expansion obtained using the v-ZSISA-QHA method (dashed lines)
with the proposed orthorhombic approach (solid lines with markers)
for three external pressures: 0, 4, and 8 GPa. Experimental data at zero pressure are 
shown as discrete points for validation. In the v-ZSISA-QHA method, thermal expansion 
is derived from calculations at eight different volumes, ranging from 0.94 to 1.08 
times the equilibrium BO volume, with increments of 0.02. The data are fitted using 
the Vinet equation of state~\cite{Vinet1987}.

The results indicate that the temperature dependence of thermal expansion predicted
by v-ZSISA-QHA and the orthorhombic ${\textrm{v-ZSISA-E}{\infty} \textrm{Vib}2}$ 
method are in good agreement, with minor discrepancies attributed to numerical errors. 
However, neither theoretical approach fully matches experimental values due to 
anharmonic effects in MgO, which are not captured within QHA. 
Nonetheless, since the objective of this study is to evaluate QHA-based methods,
the results remain valid in temperature ranges where QHA is applicable.

The bottom panel illustrates the temperature dependence of the elastic constants  $C_{11}$, 
$C_{12}$, and $C_{44}$  for $P_{\textrm{ext}}$=0 GPa (solid lines with markers), 4 GPa 
(dashed lines with markers), and 8 GPa (dotted lines with markers). Experimental 
values at $P_{\textrm{ext}}$=0 GPa are included for comparison. The 
computed elastic constants exhibit a decreasing trend with temperature and an 
increasing trend with pressure, consistent with experimental data. However,
even at 0 K, the discrepancy between theoretical elastic constants and experimental ones (at 0 GPa) is on the order of a few percent, due to the exchange-correlation functional inaccuracy.

\subsection{ZnO}
For ZnO in the wurtzite structure, the optimized lattice parameters are a=3.227 \AA, and c=5.206 \AA.
The planewave kinetic energy cutoff (ecut) is set to 42 Ha, and a k-grid and q-grid of $6\times6\times4$ is used 
for Brillouin zone sampling. For all calculations, the internal degrees of freedom of the atomic 
positions are fully optimized to minimize forces on the atoms.  
ZnO in the wurtzite structure can be analyzed using either hexagonal symmetry or
lower-symmetry configurations involving three or more degrees of freedom.
For hexagonal symmetry with three DOF, the relevant strain components are 
 $\varepsilon_{xx}$,  $\varepsilon_{yy}$, and  $\varepsilon_{zz}$,
 analogous to the orthorhombic case. However, in the hexagonal system, symmetry constraints reduce
 the independent strain components, with $\varepsilon_{xx}=\varepsilon_{yy}$ simplifying the equations to 
 a 2DOF model.

The 2DOF approach offers substantial computational advantages for two main reasons. 
First, it reduces the number of required deformations from 7 to 6, streamlining the 
overall calculation process. More importantly, it preserves hexagonal symmetry in all 
deformations, whereas the 3DOF approach breaks this symmetry, resulting in 
lower-symmetry configurations that demand more computational resources.
Since phonon spectra calculations are significantly faster for higher-symmetry structures, 
adopting the 2DOF treatment not only reduces the number of deformations, but also lowers
the computational cost for each deformation. 

Different strain configurations are required for thermal expansion and elastic constant calculations 
to account for the symmetry constraints involved in each process.
The strain patterns used for thermal expansion are
$[\varepsilon^{\textrm{BO}\bullet}]$,
$([\varepsilon^{\textrm{BO}\bullet}] +\delta \varepsilon^{\textrm{BO}}_{xx} +\delta \varepsilon^{\textrm{BO}}_{yy})$,
$([\varepsilon^{\textrm{BO}\bullet}] -\delta \varepsilon^{\textrm{BO}}_{xx} -\delta \varepsilon^{\textrm{BO}}_{yy})$,
$([\varepsilon^{\textrm{BO}\bullet}] \pm \delta \varepsilon^{\textrm{BO}}_{zz})$,
$([\varepsilon^{\textrm{BO}\bullet}] -\delta \varepsilon^{\textrm{BO}}_{xx}-\delta \varepsilon^{\textrm{BO}}_{yy} -\delta \varepsilon^{\textrm{BO}}_{zz})$.

For combined elastic constant and thermal expansion calculations, the following strain patterns are applied 
$[\varepsilon^{\textrm{BO}\bullet}]$,
$([\varepsilon^{\textrm{BO}\bullet}] \pm \delta \varepsilon^{\textrm{BO}}_{xx})$,
$([\varepsilon^{\textrm{BO}\bullet}] -\delta \varepsilon^{\textrm{BO}}_{xx} -\delta \varepsilon^{\textrm{BO}}_{yy})$,
$([\varepsilon^{\textrm{BO}\bullet}] \pm \delta \varepsilon^{\textrm{BO}}_{zz})$,
$([\varepsilon^{\textrm{BO}\bullet}] -\delta \varepsilon^{\textrm{BO}}_{xx} -\delta \varepsilon^{\textrm{BO}}_{zz})$,
$([\varepsilon^{\textrm{BO}\bullet}] + \delta \varepsilon^{\textrm{BO}}_{xz})$,
and
$([\varepsilon^{\textrm{BO}\bullet}] + 2\delta \varepsilon^{\textrm{BO}}_{xz})$ 
where the magnitude of strain increment is set to $\delta \varepsilon^{\textrm{BO}}=0.005$.
   \begin{figure}[!]
\includegraphics[width=0.4\textwidth]{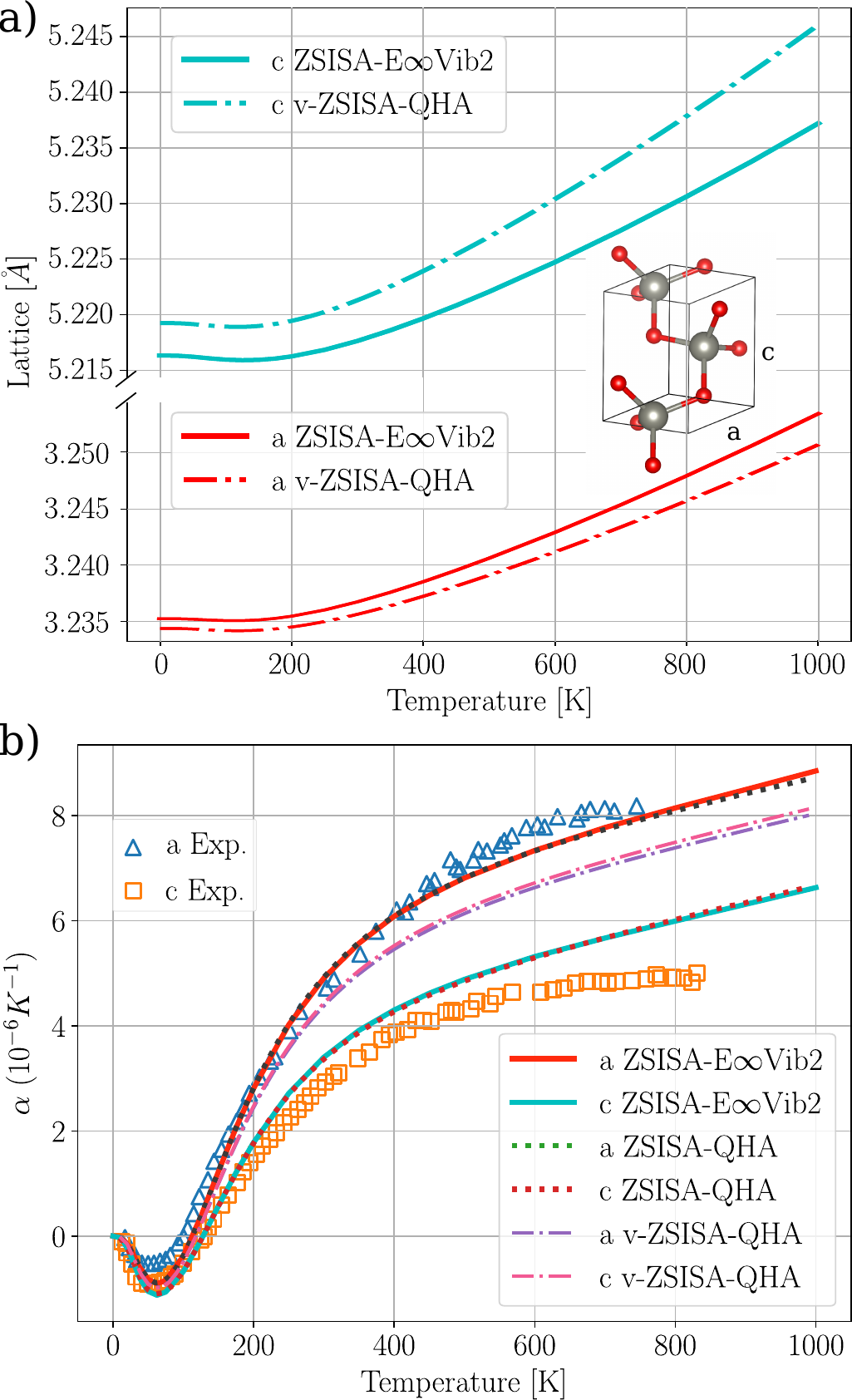}
  \caption{a) Temperature dependence of the lattice parameters a and c of ZnO, calculated 
  using the v-ZSISA-QHA approach (solid lines) and the ${\textrm{v-ZSISA-E}{\infty} \textrm{Vib}2}$ 
  method (dashed-dotted lines). b) Thermal expansion coefficients
  of a and c axes ($\alpha_a$ and $\alpha_c$)
  as a function of temperature, 
  obtained from ZSISA-QHA (dotted line), ${\textrm{v-ZSISA-E}{\infty} \textrm{Vib}2}$  (solid line),
  and v-ZSISA-QHA (dashed line). Experimental data~\cite{Ibach1969} points at zero pressure are included for validation.}
  \label{fig:ZnO1}
\end{figure}
To systematically validate the new method, we compare its results with those of 
ZSISA-QHA and v-ZSISA-QHA.
In the v-ZSISA-QHA approach, thermal expansion is determined from calculations at
seven distinct volumes, ranging from 0.96 to 1.08 times the equilibrium BO volume,
in increments of 0.02. 
For ZSISA-QHA, the lattice parameters a and c are systematically varied over a 
25-point strain mesh (five points in each direction)
spanning [-0.005,0.015] 
relative to the BO lattice. A third-order (cubic) two-dimensional surface is
then fitted to the computed free energy values, and the equilibrium lattice
parameters at each temperature are obtained by minimizing this surface.

Figure~\ref{fig:ZnO1} illustrates the temperature dependence of the lattice parameters 
and thermal expansion. The results show strong agreement between ZSISA-QHA and
${\textrm{v-ZSISA-E}{\infty} \textrm{Vib}2}$, confirming the consistency of the methodology. 
However, ZSISA-QHA and 
v-ZSISA-QHA yield different final results of anisotropic thermal expansion and lattice parameters,
while their volumetric thermal expansion predictions remain consistent. Experimental data indicate distinct 
thermal expansion coefficients for a and c, whereas v-ZSISA-QHA predicts
nearly identical values, failing to capture the experimental anisotropy.
In contrast, ${\textrm{v-ZSISA-E}{\infty} \textrm{Vib}2}$  matches experimental observations, 
effectively reproducing the anisotropic thermal expansion of ZnO.
Nevertheless, the accuracy of the results remains sensitive to the choice of the exchange-correlation
functional. 
Volumetric thermal expansion results are in agreement in v-ZSISA-QHA and ZSISA-QHA methods.

   \begin{figure}[!]
\includegraphics[width=0.43\textwidth]{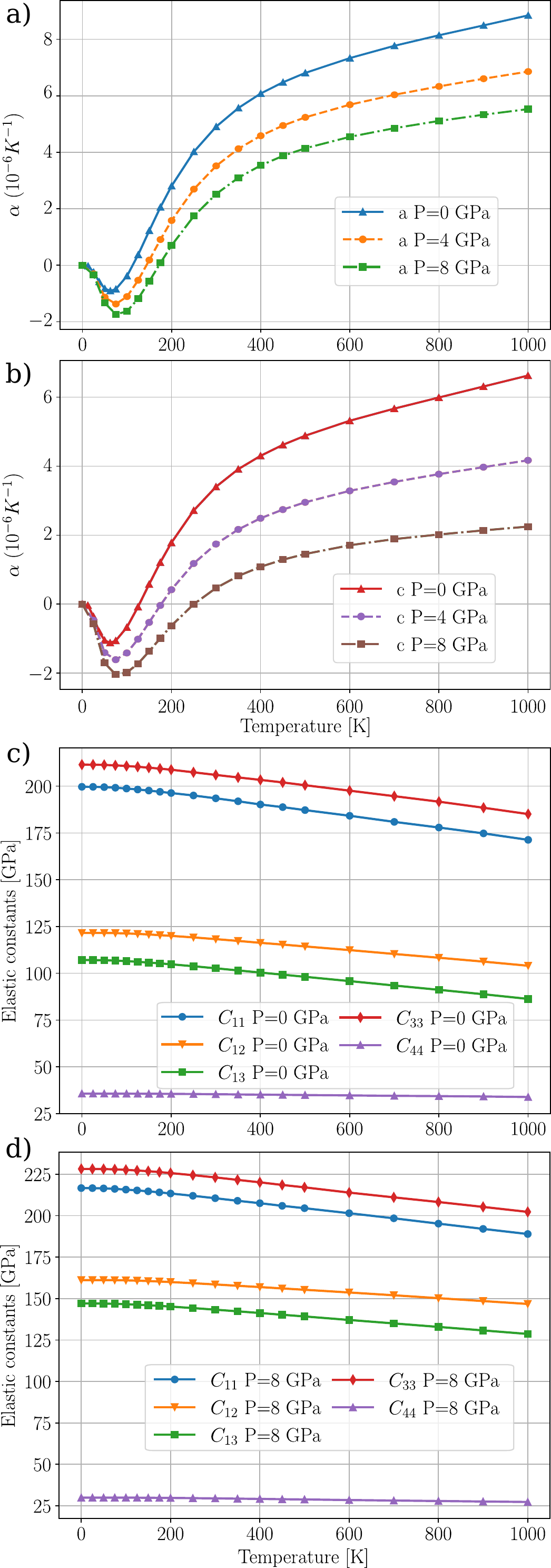}
		
  \caption{
  Temperature dependence of the anisotropic thermal expansion and elastic constants of ZnO under varying pressure conditions, as obtained from the ${\textrm{v-ZSISA-E}{\infty} \textrm{Vib}2}$ method. (a) Thermal expansion of lattice parameter a and (b) lattice parameter c at pressures of 0, 4, and 8 GPa. (c) Elastic constants at 0 GPa and (d) elastic constants at 8 GPa.}
  \label{fig:ZnO2}
\end{figure}

 Figure~\ref{fig:ZnO2} presents the temperature dependence of the anisotropic thermal 
 expansion and elastic constants of ZnO under varying pressure conditions, as obtained 
 from the  ${\textrm{v-ZSISA-E}{\infty} \textrm{Vib}2}$ method. Panels (a) and (b) show
 the thermal expansion of the lattice parameters a and c at pressures of 0, 4, and 8 
 GPa, respectively. Panels (c) and (d) display the corresponding elastic constants at 0 
 GPa and 8 GPa, highlighting the effect of pressure on the mechanical properties of 
 ZnO. 
\subsection{\ce{ZrO2}}

\ce{ZrO2} adopts a monoclinic structure at temperatures below  approximately 1443 K.
In our calculations, the optimized lattice parameters are a=5.126 \AA, b=5.209 \AA, c=5.295 \AA,
with a monoclinic angle $\beta=99.59$. 
The strain configurations applied for the monoclinic case in determining both thermal 
expansion and elastic constants differ by only three additional strains required for 
elastic constants. Specifically, 15 strain patterns are sufficient for thermal 
expansion calculations, while 18 strains are needed when computing elastic constants. 
The complete set of these strain configurations is provided in Table \ref{tab:strain} in the appendix.
 \begin{figure}[t!]
\includegraphics[width=0.44\textwidth]{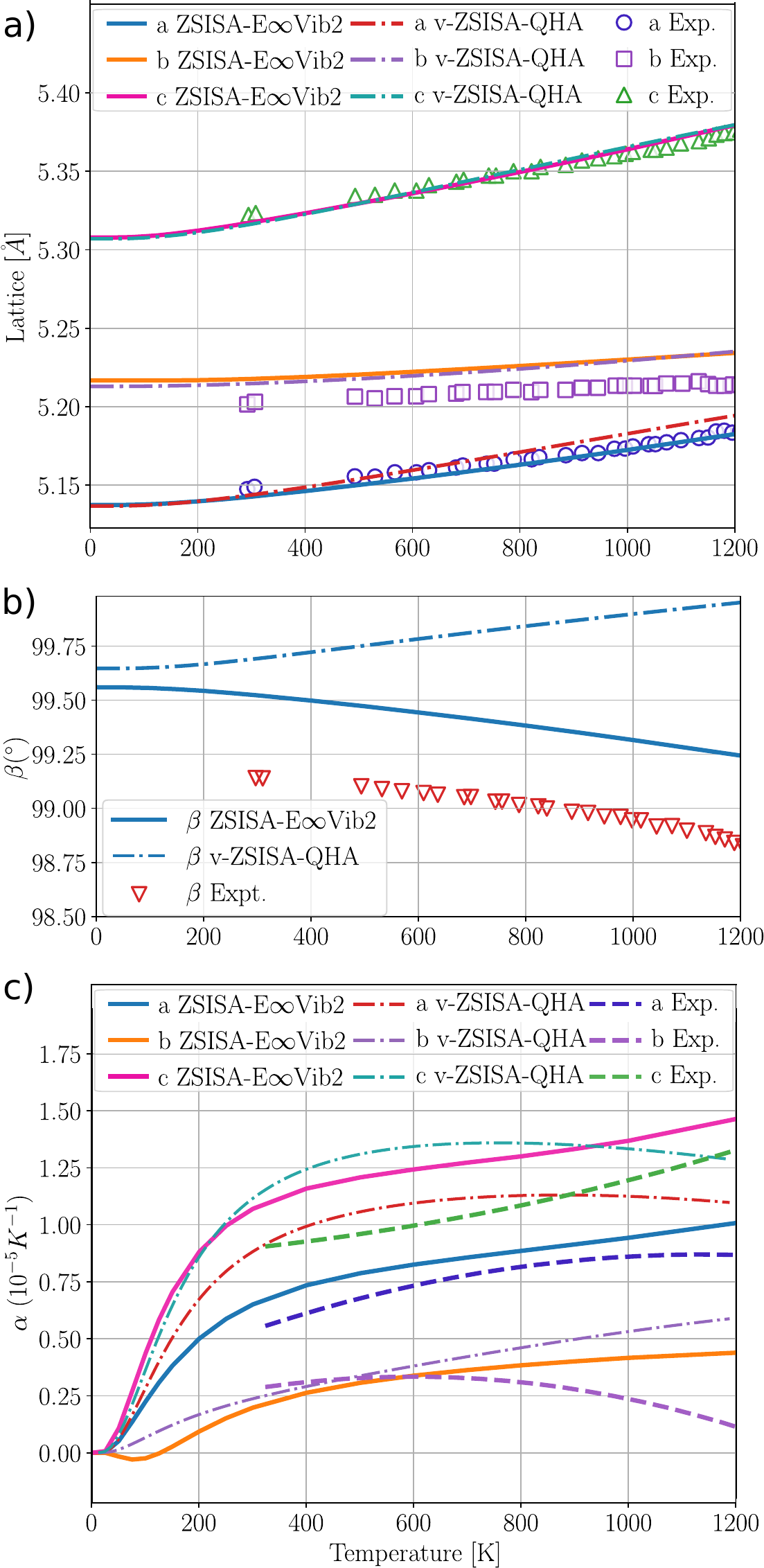}
  \caption{Temperature dependence of (a) the lattice parameters and (b) the angle of \ce{ZrO2}, 
  calculated using the ${\textrm{v-ZSISA-E}{\infty} \textrm{Vib}2}$  
  approach (solid lines) and the  v-ZSISA-QHA method (dashed-dotted lines). 
  Experimental data~\cite{Haggerty2014} points are shown as markers. (c) Thermal expansion coefficient as 
  a function of temperature, with results from the  ${\textrm{v-ZSISA-E}{\infty} 
  \textrm{Vib}2}$  method (solid line) and the v-ZSISA-QHA approach (dashed-dotted 
  line). Experimental thermal expansion data are provided by the dashed lines for 
  comparison.}
  \label{fig:ZrO2}
\end{figure}

Figure~\ref{fig:ZrO2} illustrates the temperature dependence of the lattice parameters and
angle of monoclinic \ce{ZrO2} at zero pressure. Panels (a) and (b) show the lattice parameters 
and the angle $\beta$, respectively, calculated using the ${\textrm{v-ZSISA-E}{\infty}   \textrm{Vib}2}$ 
approach (solid lines) and the v-ZSISA-QHA   method (dashed-dotted lines). Experimental data points 
are included for direct comparison. Panel (c) presents the thermal expansion coefficient as a function 
of temperature, with results from the  ${\textrm{v-ZSISA-E}{\infty}   \textrm{Vib}2}$ method 
(solid line) and the v-ZSISA-QHA approach (dashed-dotted line), with the experimental thermal 
expansion data shown by the dashed lines for reference.

The v-ZSISA-QHA method is based on calculations at seven volumes, ranging from 0.96\% to 1.08\% 
of the BO volume, with a step size of 0.02\%. The experimental thermal expansion is computed by 
fitting a line to the lattice data points at different temperatures and differentiating the fitted curve.

A noteworthy difference between the v-ZSISA-QHA and  ${\textrm{ZSISA-E}{\infty}   \textrm{Vib}2}$ 
methods is the temperature dependence of the angle. Experimental results~\cite{Haggerty2014} show a decrease in the $\beta$ 
angle with increasing temperature. The v-ZSISA-QHA 
method predicts an opposite trend, where the angle increases with temperature. In contrast, the 
${\textrm{v-ZSISA-E}{\infty}   \textrm{Vib}2}$ method correctly captures the direction of the angle
decrease, although the predicted values are slightly different. This discrepancy is due to the exchange-correlation functional. Modifying the functional
could potentially lead to a better agreement with experimental results. The thermal expansion 
predictions from the  ${\textrm{v-ZSISA-E}{\infty}   \textrm{Vib}2}$ show a better agreement 
with the experimental data compared to the v-ZSISA-QHA approach.

 \begin{figure}[t!]
\includegraphics[width=0.48\textwidth]{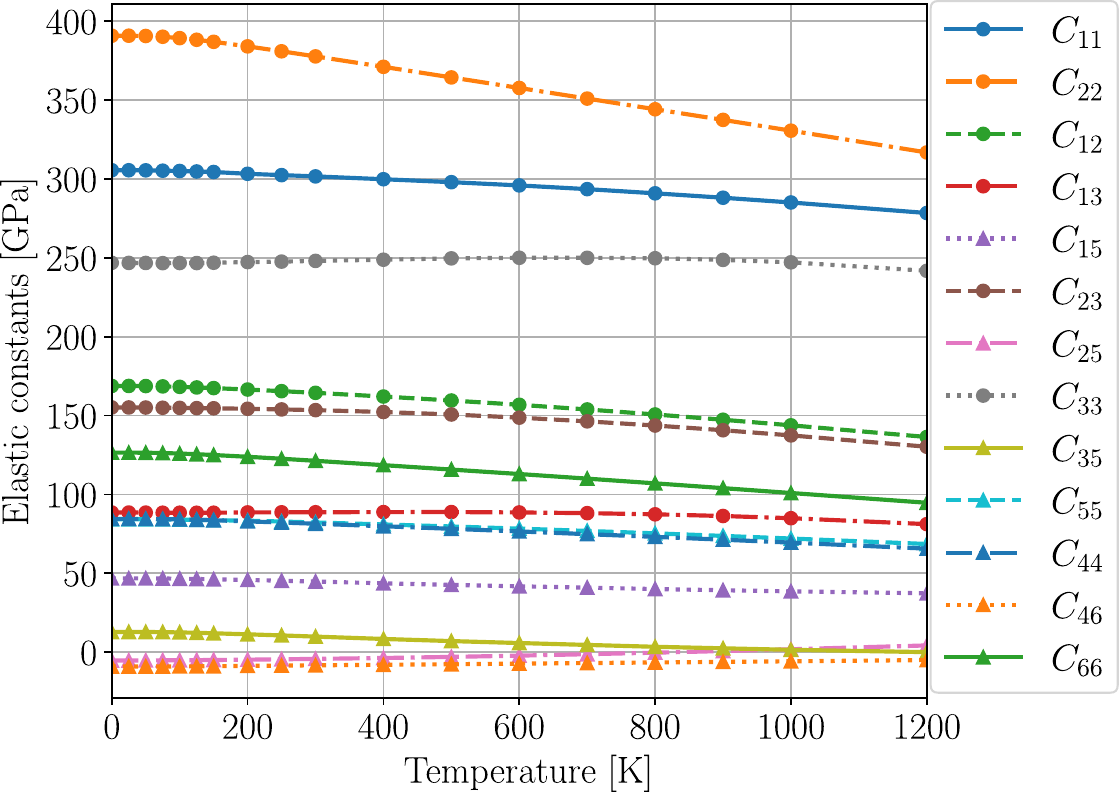}
  \caption{Temperature dependence of the elastic constants of monoclinic \ce{ZrO2}}
  \label{fig:ZrO2_elastic}
\end{figure}

Figure~\ref{fig:ZrO2_elastic} illustrates the temperature dependence of the 13 elastic constants of monoclinic structure, emphasizing their temperature-induced variations.

\section{CONCLUSION}\label{sec:conclusion}
In this work, we introduce a novel method for determining the anisotropic thermal 
expansion and elastic constants of materials with arbitrary crystal structures, achieving accuracy comparable to the quasiharmonic approximation with zero static 
internal stress (ZSISA-QHA). While ZSISA-QHA is highly accurate for weakly anharmonic crystals, it is computationally 
expensive due to the need for numerous phonon spectrum calculations. Specifically, for
systems with $n$ lattice degrees of freedom, traditional methods require an impractically 
large number of lattice mesh points, often exceeding $5^n$ for high-accuracy thermal expansion predictions. This makes the direct application of ZSISA-QHA computationally intensive for systems with more than three lattice degrees of freedom. 
Our method addresses this limitation by reducing the number of required calculations, making it applicable to a wider range of systems while preserving a sufficient precision up to about 800 K.

Building on prior work~\cite{Rostami2024}, where we demonstrated that truncating the vibrational free 
energy expansion to second order provides results comparable to the full quasiharmonic 
treatment, we extend this approach by employing a Taylor series expansion of the 
vibrational free energy up to the second derivative to calculate the Gibbs free energy 
in the multidimensional space of degrees of freedom. This method incorporates self-consistent optimization of 
lattice parameters and atomic positions, taking into account thermal stresses and using
the accurate Born-Oppenheimer energy to ensure precise thermal properties.

By significantly reducing the computational demands, our approach provides a practical 
and efficient alternative to ZSISA-QHA, especially for materials with complex crystal 
symmetries, such as monoclinic and triclinic structures. For instance, in the case of a
triclinic system, we achieve accurate predictions of both thermal expansion and elastic
constants with only 28 phonon spectrum calculations—three order of magnitude reduction 
compared to more than 15625 calculations required by ZSISA-QHA.

Our results demonstrate that the new method can replicate ZSISA-QHA with high accuracy,
making it a viable alternative whenever ZSISA-QHA is not applicable. We successfully apply 
this method to 10 materials with a variety of crystal structures, from cubic to 
monoclinic and triclinic forms, further validating its utility and versatility.

\begin{acknowledgments} 
This work has been supported by the Fonds de la Recherche
Scientifique (FRS-FNRS Belgium) through the PdR Grant No.
T.0103.19 – ALPS. 
It is an outcome of the Shapeable
2D magnetoelectronics by design project (SHAPEme, EOS
Project No. 560400077525) that has received funding from
the FWO and FRS-FNRS under the Belgian Excellence of
Science (EOS) program.
Computational resources have been provided by the supercomputing facilities
of the Universit\'e catholique de Louvain (CISM/UCL) and the Consortium des
Equipements de Calcul Intensif en F\'ed\'eration Wallonie Bruxelles (CECI) 
funded by the FRS-FNRS under Grant No. 2.5020.11.
\end{acknowledgments}

\section{Appendix}\label{sec:appendix}

In this appendix, we present the equations relevant for different crystallographic systems,
namely the cubic system, uniaxial system, triclinic system, isotropic slab, anisotropic slab (two degrees of freedom), and anisotropic slab.
They correspond to the equations presented for the orthorhombic and monoclinic systems in the body of the paper, namely 
Eqs.(\ref{eq:R_orthorhombic}) to (\ref{eq:Stress_orthorhombic}) for the orthorhombic system, and Eqs.(\ref{eq:R-monoclinic}) to (\ref{eq:stress_monoclinic}) for the
monoclinic system.

For the sake of compactness, in this appendix, the $P_{\textrm{ext}}$ dependence of the components of
the lattice vectors is not explicitly indicated, unlike their
temperature dependence. 

\subsection{Cubic}
  \begin{align}
		[R]&=	 \left[ \begin{array}{lll}
		(1+\varepsilon^{\textrm{BO}}_{xx}) R^{\textrm{BO}}_{1,x}
		&(1+\varepsilon^{\textrm{BO}}_{xx}) R^{\textrm{BO}}_{2,x}
		&  (1+\varepsilon^{\textrm{BO}}_{xx})  R^{\textrm{BO}}_{3,x}  \\
		(1+\varepsilon^{\textrm{BO}}_{xx})  R^{\textrm{BO}}_{1,y}
		&  (1+\varepsilon^{\textrm{BO}}_{xx}) R^{\textrm{BO}}_{2,y}
		&  (1+\varepsilon^{\textrm{BO}}_{xx})  R^{\textrm{BO}}_{3,y}  \\
		(1+\varepsilon^{\textrm{BO}}_{xx}) R^{\textrm{BO}}_{1,z}
		& (1+\varepsilon^{\textrm{BO}}_{xx}) R^{\textrm{BO}}_{2,z}
		& (1+\varepsilon^{\textrm{BO}}_{xx})  R^{\textrm{BO}}_{3,z} 
	\end{array} \right]
 \end{align} 
\begin{align}
		\varepsilon^{\textrm{BO}}_{xx}(T)= \frac{A_x(T)}{A^{\textrm{BO}}_x}-1 \quad,\quad \varepsilon^{\textrm{BO}\bullet}_{xx}= \frac{A^{\bullet}_x}{A^{\textrm{BO}}_x}-1.
	\end{align}
\begin{align}\nonumber
 \sigma^{\textrm{vib}}_{xx}\Big|_{[R(T)]}=& \frac{1}{3}   \frac{1}{V(T)}\frac{dF_{\textrm{vib}}}{d\varepsilon^{\textrm{BO}}_{xx}}\Big|_{[R(T)],T}\left(1+\varepsilon^{\textrm{BO}}_{xx}(T)\right).\\
 \sigma^{\textrm{vib}}_{yy}\Big|_{[R(T)]}=&\sigma^{\textrm{vib}}_{zz}\Big|_{[R(T)]} =\sigma^{\textrm{vib}}_{xx}\Big|_{[R(T)]}.
  \end{align}
  \\
\subsection{Hexagonal, Trigonal,  Tetragonal}
  \begin{align}
		[R]&=	 \left[ \begin{array}{lll}
		(1+\varepsilon^{\textrm{BO}}_{xx}) R^{\textrm{BO}}_{1,x}
		&(1+\varepsilon^{\textrm{BO}}_{xx}) R^{\textrm{BO}}_{2,x}
		&  (1+\varepsilon^{\textrm{BO}}_{xx})  R^{\textrm{BO}}_{3,x}  \\
		(1+\varepsilon^{\textrm{BO}}_{xx})  R^{\textrm{BO}}_{1,y}
		&  (1+\varepsilon^{\textrm{BO}}_{xx}) R^{\textrm{BO}}_{2,y}
		&  (1+\varepsilon^{\textrm{BO}}_{xx})  R^{\textrm{BO}}_{3,y}  \\
		(1+\varepsilon^{\textrm{BO}}_{zz}) R^{\textrm{BO}}_{1,z}
		& (1+\varepsilon^{\textrm{BO}}_{zz}) R^{\textrm{BO}}_{2,z}
		& (1+\varepsilon^{\textrm{BO}}_{zz})  R^{\textrm{BO}}_{3,z} 
	\end{array} \right]
 \end{align} 
\begin{align}\nonumber
		\varepsilon^{\textrm{BO}}_{xx}(T)= \frac{A_x(T)}{A^{\textrm{BO}}_x}-1 \quad,\quad \varepsilon^{\textrm{BO}\bullet}_{xx}= \frac{A^{\bullet}_x}{A^{\textrm{BO}}_x}-1.\\
		\varepsilon^{\textrm{BO}}_{zz}(T)= \frac{C_z(T)}{C^{\textrm{BO}}_z}-1 \quad,\quad \varepsilon^{\textrm{BO}\bullet}_{zz}= \frac{C^{\bullet}_z}{C^{\textrm{BO}}_z}-1.
	\end{align}
\begin{align}\nonumber
 \sigma^{\textrm{vib}}_{xx}\Big|_{[R(T)]}=& \frac{1}{2}  \frac{1}{V(T)}\frac{dF_{\textrm{vib}}}{d\varepsilon^{\textrm{BO}}_{xx}}\Big|_{[R(T)],T}\left(1+\varepsilon^{\textrm{BO}}_{xx}(T)\right)\\
		\sigma^{\textrm{vib}}_{yy}\Big|_{[R(T)]}=&   \sigma^{\textrm{vib}}_{xx}\Big|_{[R(T)]}\\
		\sigma^{\textrm{vib}}_{zz}\Big|_{[R(T)]}=&   \frac{1}{V(T)}\frac{dF_{\textrm{vib}}}{d\varepsilon^{\textrm{BO}}_{zz}}\Big|_{[R(T)],T}\left(1+\varepsilon^{\textrm{BO}}_{zz}(T)\right)\nonumber
  \end{align}
\\

\subsection{Triclinic}
    \begin{align}
		&[R]= 
	\left[ \begin{array}{lll}
	R_{1,x} & 	R_{2,x}  & 	R_{3,x}  \\ 
		0& 	R_{2,y} & 	R_{3,y} \\
			0 & 	0 & 	R_{3,z}  
\end{array} \right]\\
&~~~~=
\left[ \begin{array}{lll}
	1+\varepsilon^{\textrm{BO}}_{xx} & 	\varepsilon^{\textrm{BO}}_{xy}& 	\varepsilon^{\textrm{BO}}_{xz}   \\ 
		0 & 	1+\varepsilon^{\textrm{BO}}_{yy} & 	\varepsilon^{\textrm{BO}}_{yz} \\
	0& 	0 & 	1+\varepsilon^{\textrm{BO}}_{zz}
\end{array} \right]
[R^{\textrm{BO}}].
\nonumber
	\end{align}

   \begin{widetext}

  \begin{align}\nonumber
		[R]&=	 \left[ \begin{array}{lll}
		(1+\varepsilon^{\textrm{BO}}_{xx}) R^{\textrm{BO}}_{1,x}
		&  (1+\varepsilon^{\textrm{BO}}_{xx})  R^{\textrm{BO}}_{2,x}+   \varepsilon^{\textrm{BO}}_{xy}   R^{\textrm{BO}}_{2,y}
		&  (1+\varepsilon^{\textrm{BO}}_{xx}) R^{\textrm{BO}}_{3,x}+   \varepsilon^{\textrm{BO}}_{xy}  R^{\textrm{BO}}_{3,y}+   \varepsilon^{\textrm{BO}}_{xz}    R^{\textrm{BO}}_{3,z}  \\
		0
		& (1+\varepsilon^{\textrm{BO}}_{yy}) R^{\textrm{BO}}_{2,y}
		&  (1+\varepsilon^{\textrm{BO}}_{yy}) R^{\textrm{BO}}_{3,y}+   \varepsilon^{\textrm{BO}}_{yz}   R^{\textrm{BO}}_{3,z} \\
		0
		&  0
		&  (1+\varepsilon^{\textrm{BO}}_{zz})  R^{\textrm{BO}}_{3,z} 
	\end{array} \right]
	\end{align} 

\begin{align}\nonumber
		\varepsilon^{\textrm{BO}}_{xx}(T)=& \frac{R_{1,x}(T)}{R_{1,x}^{\textrm{BO}}}-1\quad,\quad \varepsilon^{\textrm{BO}\bullet}_{xx}= \frac{R_{1,x}^{\bullet}}{R_{1,x}^{\textrm{BO}}}-1.\\
		\varepsilon^{\textrm{BO}}_{yy}(T)=& \frac{R_{2,y}(T)}{R_{2,y}^{\textrm{BO}}}-1 \quad,\quad \varepsilon^{\textrm{BO}\bullet}_{yy}= \frac{R_{2,y}^{\bullet}}{R_{2,y}^{\textrm{BO}}}-1.\\
		\varepsilon^{\textrm{BO}}_{zz}(T)=& \frac{R_{3,z}(T)}{R_{3,z}^{\textrm{BO}}}-1 \quad,\quad \varepsilon^{\textrm{BO}\bullet}_{zz}= \frac{R_{3,z}^{\bullet}}{R_{3,z}^{\textrm{BO}}}-1.\nonumber
  \end{align}
\begin{align*}
  \varepsilon^{\textrm{BO}}_{xy}(T)=& \frac{R_{2,x}(T) R_{1,x}^{\textrm{BO}}-R_{1,x}(T) R_{2,x}^{\textrm{BO}}}{R_{1,x}^{\textrm{BO}} R_{2,y}^{\textrm{BO}}}\quad,\quad
  \varepsilon^{\textrm{BO}\bullet}_{xy}=  \frac{R_{2,x}^{\bullet} R_{1,x}^{\textrm{BO}}-R_{1,x}^{\bullet} R_{2,x}^{\textrm{BO}}}{R_{1,x}^{\textrm{BO}} R_{2,y}^{\textrm{BO}}}.\\
   \varepsilon^{\textrm{BO}}_{yz}(T)=& \frac{R_{3,y}(T) R_{2,y}^{\textrm{BO}}-R_{2,y}(T) R_{3,y}^{\textrm{BO}}}{R_{2,y}^{\textrm{BO}} R_{3,z}^{\textrm{BO}}}\quad,\quad
  \varepsilon^{\textrm{BO}\bullet}_{yz}= \frac{R_{3,y}^{\bullet} R_{2,y}^{\textrm{BO}}-R_{2,y}^{\bullet} R_{3,y}^{\textrm{BO}}}{R_{2,y}^{\textrm{BO}} R_{3,z}^{\textrm{BO}}}.
	\end{align*}

 \begin{align*}
  \varepsilon^{\textrm{BO}}_{xz}(T)=&\frac{R_{1,x}^{\textrm{BO}} (R_{2,y}^{\textrm{BO}}R_{3,x}(T) - R_{2,x}(T)  R_{3,y}^{\textrm{BO}})-R_{1,x}(T) (R_{2,y}^{\textrm{BO}} R_{3,x}^{\textrm{BO}} - 
  R_{2,x}^{\textrm{BO}} R_{3,y}^{\textrm{BO}})}{R_{1,x}^{\textrm{BO}} R_{2,y}^{\textrm{BO}} R_{3,z}^{\textrm{BO}}}.\\
   \varepsilon^{\textrm{BO}\bullet}_{xz}=&\frac{R_{1,x}^{\textrm{BO}} (R_{2,y}^{\textrm{BO}}R_{3,x}^{\bullet} - R_{2,x}^{\bullet}  R_{3,y}^{\textrm{BO}})-R_{1,x}^{\bullet} (R_{2,y}^{\textrm{BO}} R_{3,x}^{\textrm{BO}} - 
  R_{2,x}^{\textrm{BO}} R_{3,y}^{\textrm{BO}})}{R_{1,x}^{\textrm{BO}} R_{2,y}^{\textrm{BO}} R_{3,z}^{\textrm{BO}}}.
	\end{align*}
    
 \begin{align}\nonumber
\sigma^{\textrm{vib}}_{xx}\Big|_{[R(T)]}=&  \frac{1}{V(T)}\frac{dF_{\textrm{vib}}}{d\varepsilon^{\textrm{BO}}_{xx}}\Big|_{[R(T)],T}\left(1+\varepsilon^{\textrm{BO}}_{xx}(T)\right).\\\nonumber
	\sigma^{\textrm{vib}}_{yy}\Big|_{[R(T)]}=&  \frac{1}{V(T)}\frac{dF_{\textrm{vib}}}{d\varepsilon^{\textrm{BO}}_{yy}}\Big|_{[R(T)],T}\left(1+\varepsilon^{\textrm{BO}}_{yy}(T)\right).\\
	\sigma^{\textrm{vib}}_{zz}\Big|_{[R(T)]}=&   \frac{1}{V(T)}\frac{dF_{\textrm{vib}}}{d\varepsilon^{\textrm{BO}}_{zz}}\Big|_{[R(T)],T}\left(1+\varepsilon^{\textrm{BO}}_{zz}(T)\right).\\\nonumber
 \sigma^{\textrm{vib}}_{xy}\Big|_{[R(T)]}=&  \frac{1}{V(T)}
  \left(\frac{dF_{\textrm{vib}}}{d\varepsilon^{\textrm{BO}}_{xy}}\Big|_{[R(T)],T}\left(1+\varepsilon^{\textrm{BO}}_{yy}(T)\right)
  + \frac{dF_{\textrm{vib}}}{d\varepsilon^{\textrm{BO}}_{xx}}\Big|_{[R(T)],T} \varepsilon^{\textrm{BO}}_{xy}(T)\right).\\\nonumber
    \sigma^{\textrm{vib}}_{yz}\Big|_{[R(T)]}=&  \frac{1}{V(T)}
  \left(\frac{dF_{\textrm{vib}}}{d\varepsilon^{\textrm{BO}}_{yz}}\Big|_{[R(T)],T}\left(1+\varepsilon^{\textrm{BO}}_{zz}(T)\right)
  + \frac{dF_{\textrm{vib}}}{d\varepsilon^{\textrm{BO}}_{yy}}\Big|_{[R(T)],T} \varepsilon^{\textrm{BO}}_{yz}(T)\right).\\
   \sigma^{\textrm{vib}}_{xz}\Big|_{[R(T)]}=& 
  \frac{1}{V(T)}
  \left(\frac{dF_{\textrm{vib}}}{d\varepsilon^{\textrm{BO}}_{xz}}\Big|_{[R(T)],T}\left(1+\varepsilon^{\textrm{BO}}_{zz}(T)\right)
  + \frac{dF_{\textrm{vib}}}{d\varepsilon^{\textrm{BO}}_{xy}}\Big|_{[R(T)],T} \varepsilon^{\textrm{BO}}_{yz}(T)  
  + \frac{dF_{\textrm{vib}}}{d\varepsilon^{\textrm{BO}}_{xx}}\Big|_{[R(T)],T}\varepsilon^{\textrm{BO}}_{xz}(T)  \right ). \nonumber
  \end{align}
  \end{widetext}
\subsection{Isotropic slab}
  \begin{align}
		[R]&=	 \left[ \begin{array}{lll}
		(1+\varepsilon^{\textrm{BO}}_{xx}) R^{\textrm{BO}}_{1,x}
		&(1+\varepsilon^{\textrm{BO}}_{xx}) R^{\textrm{BO}}_{1,y}
		& 0  \\
		(1+\varepsilon^{\textrm{BO}}_{xx})  R^{\textrm{BO}}_{2,x}
		&  (1+\varepsilon^{\textrm{BO}}_{xx}) R^{\textrm{BO}}_{2,y}
		&  0 \\
		0
		& 0
		&   R^{\textrm{BO}}_{3,z} 
	\end{array} \right]
 \end{align} 
\begin{align}
		\varepsilon^{\textrm{BO}}_{xx}(T)= \frac{A_x(T)}{A^{\textrm{BO}}_x}-1 \quad,\quad \varepsilon^{\textrm{BO}\bullet}_{xx}= \frac{A^{\bullet}_x}{A^{\textrm{BO}}_x}-1.
	\end{align}
\begin{align}\nonumber
 \sigma^{\textrm{vib}}_{xx}\Big|_{[R(T)]}=& \frac{1}{2}   \frac{1}{V(T)}\frac{dF_{\textrm{vib}}}{d\varepsilon^{\textrm{BO}}_{xx}}\Big|_{[R(T)],T}\left(1+\varepsilon^{\textrm{BO}}_{xx}(T)\right).\\
 \sigma^{\textrm{vib}}_{yy}\Big|_{[R(T)]}=& \sigma^{\textrm{vib}}_{xx}\Big|_{[R(T)]}.
  \end{align}

\subsection{Anisotropic slab (2DOF)}
  \begin{align}
		[R]&=	 \left[ \begin{array}{lll}
		(1+\varepsilon^{\textrm{BO}}_{xx}) R^{\textrm{BO}}_{1,x}
		&(1+\varepsilon^{\textrm{BO}}_{yy}) R^{\textrm{BO}}_{1,y}
		& 0 \\
		(1+\varepsilon^{\textrm{BO}}_{xx})  R^{\textrm{BO}}_{2,x}
		&  (1+\varepsilon^{\textrm{BO}}_{yy}) R^{\textrm{BO}}_{2,y}
		& 0 \\
		0
		& 0
		&   R^{\textrm{BO}}_{3,z} 
	\end{array} \right]
 \end{align} 
\begin{align}\nonumber
		\varepsilon^{\textrm{BO}}_{xx}(T)= \frac{A_x(T)}{A^{\textrm{BO}}_x}-1 \quad,\quad \varepsilon^{\textrm{BO}\bullet}_{xx}= \frac{A^{\bullet}_x}{A^{\textrm{BO}}_x}-1.\\
		\varepsilon^{\textrm{BO}}_{yy}(T)= \frac{B_y(T)}{B^{\textrm{BO}}_y}-1 \quad,\quad \varepsilon^{\textrm{BO}\bullet}_{yy}= \frac{B^{\bullet}_y}{B^{\textrm{BO}}_y}-1.
	\end{align}
\begin{align}\nonumber
 \sigma^{\textrm{vib}}_{xx}\Big|_{[R(T)]}=&  \frac{1}{V(T)}\frac{dF_{\textrm{vib}}}{d\varepsilon^{\textrm{BO}}_{xx}}\Big|_{[R(T)],T}\left(1+\varepsilon^{\textrm{BO}}_{xx}(T)\right).\\
		\sigma^{\textrm{vib}}_{yy}\Big|_{[R(T)]}=&    \frac{1}{V(T)}\frac{dF_{\textrm{vib}}}{d\varepsilon^{\textrm{BO}}_{yy}}\Big|_{[R(T)],T}\left(1+\varepsilon^{\textrm{BO}}_{yy}(T)\right).
  \end{align}

\subsection{Anisotropic slab}
  \begin{align}
		&[R]= 
	\left[ \begin{array}{lll}
	R_{1,x} & 	R_{2,x} & 	0  \\ 
		0 & 	R_{2,y} & 	0 \\
			0 & 	0 & 	R_{3,z}  
\end{array} \right]\\
&~~~~=
\left[ \begin{array}{lll}
	1+\varepsilon^{\textrm{BO}}_{xx} & 	\varepsilon^{\textrm{BO}}_{xy}  & 	0  \\ 
		0 & 	1+\varepsilon^{\textrm{BO}}_{yy} & 	0 \\
	0& 	0 & 	1
\end{array} \right].[R^{\textrm{BO}}]=\nonumber\\ 
&\left[ \begin{array}{lll}
		(1+\varepsilon^{\textrm{BO}}_{xx}) R^{\textrm{BO}}_{1,x}
		&  (1+\varepsilon^{\textrm{BO}}_{xx}) R^{\textrm{BO}}_{2,x}+  \varepsilon^{\textrm{BO}}_{xy}    R^{\textrm{BO}}_{2,y} 
		&0 \\
		0 &  (1+\varepsilon^{\textrm{BO}}_{yy}) R^{\textrm{BO}}_{2,y}
		& 0  \\
		0
		&  0
		&   R^{\textrm{BO}}_{3,z} 
	\end{array} \right]\nonumber
	\end{align} 

\begin{align}\nonumber
		\varepsilon^{\textrm{BO}}_{xx}(T)=& \frac{R_{1,x}(T)}{R_{1,x}^{\textrm{BO}}}-1 \quad,\quad \varepsilon^{\textrm{BO}\bullet}_{xx}= \frac{R_{1,x}^{\bullet}}{R_{1,x}^{\textrm{BO}}}-1\\ \nonumber
		\varepsilon^{\textrm{BO}}_{yy}(T)=& \frac{R_{2,y}(T)}{R_{2,y}^{\textrm{BO}}}-1 \quad,\quad \varepsilon^{\textrm{BO}\bullet}_{yy}= \frac{R_{2,y}^{\bullet}}{R_{2,y}^{\textrm{BO}}}-1\\
  \varepsilon^{\textrm{BO}}_{xy}(T)=& \frac{R_{2,x}(T) R_{1,x}^{\textrm{BO}}-R_{1,x}(T) R_{2,x}^{\textrm{BO}}}{R_{1,x}^{\textrm{BO}} R_{2,y}^{\textrm{BO}}}\\
  \varepsilon^{\textrm{BO}\bullet}_{xy}=&  \frac{R_{2,x}^{\bullet} R_{1,x}^{\textrm{BO}}-R_{1,x}^{\bullet} R_{2,x}^{\textrm{BO}}}{R_{1,x}^{\textrm{BO}} R_{2,y}^{\textrm{BO}}}\nonumber
	\end{align}

\begin{align}\nonumber
& \sigma^{\textrm{vib}}_{xx}\Big|_{[R(T)]}=  \frac{1}{V(T)}\frac{dF_{\textrm{vib}}}{d\varepsilon^{\textrm{BO}}_{xx}}\Big|_{[R(T)],T}\left(1+\varepsilon^{\textrm{BO}}_{xx}(T)\right)\\
	&	\sigma^{\textrm{vib}}_{yy}\Big|_{[R(T)]}=  \frac{1}{V(T)}\frac{dF_{\textrm{vib}}}{d\varepsilon^{\textrm{BO}}_{yy}}\Big|_{[R(T)],T}\left(1+\varepsilon^{\textrm{BO}}_{yy}(T)\right)\\\nonumber
 & \sigma^{\textrm{vib}}_{xy}\Big|_{[R(T)]}=  \frac{1}{V(T)}
  \left(\frac{dF_{\textrm{vib}}}{d\varepsilon^{\textrm{BO}}_{xy}}\Big|_{[R(T)],T}\left(1+\varepsilon^{\textrm{BO}}_{yy}(T)\right)\right .
 \\
  &+\left .\frac{dF_{\textrm{vib}}}{d\varepsilon^{\textrm{BO}}_{xx}}\Big|_{[R(T)],T}
\left(\varepsilon^{\textrm{BO}}_{xy}(T)\right) \right)\nonumber
  \end{align}~\\
  
\subsection{Accurate second derivative calculation}
When selecting strain points for determining the second derivative of the vibrational free energy, special care must be taken to ensure accuracy, particularly at points where the first derivative is zero. The vibrational free energy is expected to be symmetric around zero strain. However, applying strain to high-symmetry crystal structures often reduces their space group symmetry. This effect is particularly pronounced in systems with orthogonal angles, where numerical artifacts in the code implementation can introduce small energy shifts, potentially affecting the accuracy of second derivative calculations.

\begin{figure}[!]
\includegraphics[width=0.48\textwidth]{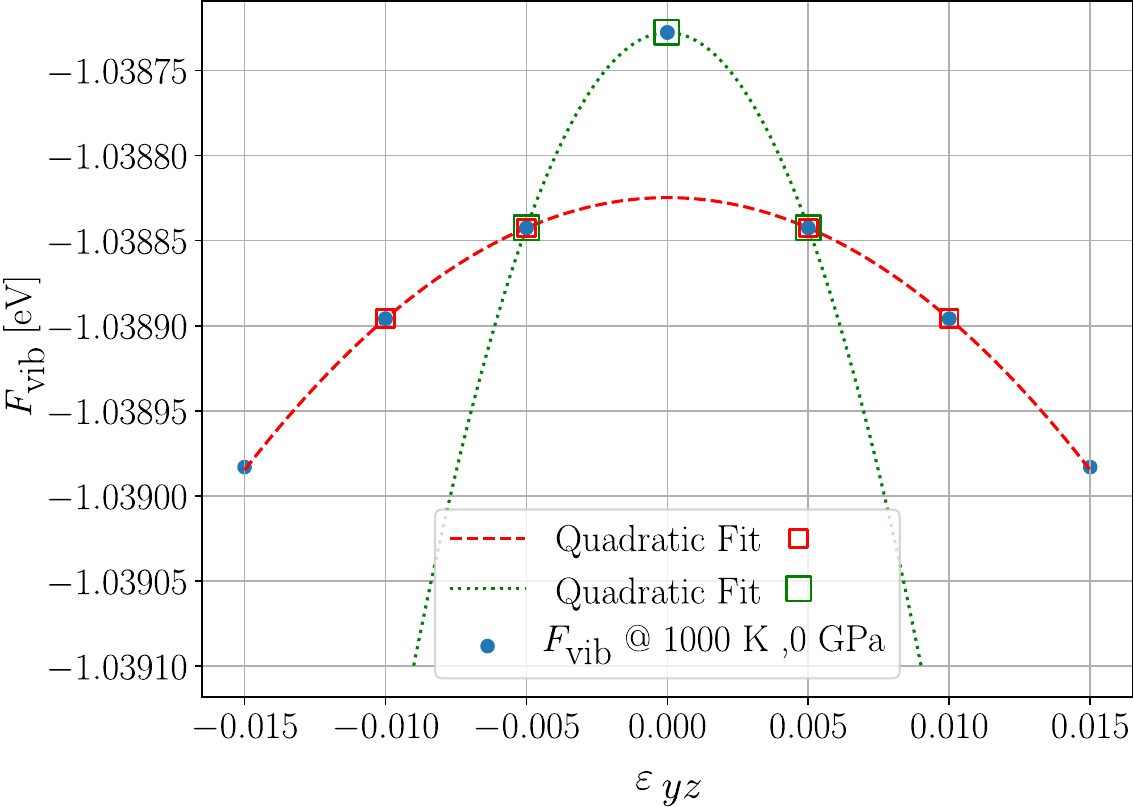}
  \caption{Vibrational free energy of ZnO as a function of strain $\varepsilon_{yz}$ at 1000 K and 0 GPa. Circles represent F$_\textrm{vib}$. The red dashed line corresponds to a quadratic fit on the reference square data points, while the green dotted line represents a quadratic fit on the green square data points. Note the smallness of the vertical scale - small energy differences are shown here.
  \label{fig:Fvib}
}
\end{figure}

As illustrated in 
Fig.~\ref{fig:Fvib} for the vibrational free energy of ZnO at 1000 K and 0 GPa, the energy follows a smooth curve, but a slight shift at zero strain is observed.
This shift is unexplained at the time of writing. If only two symmetrically placed points around zero strain ($\pm \varepsilon$) were used for fitting, the resulting quadratic curve might not fully capture the trend of the data. To correct this behavior, a quadratic curve fitting on four points ($\pm \varepsilon$) and ($\pm 2\varepsilon$ ) can be employed, ensuring a more representative fit that accounts for such numerical fluctuation. Due to symmetry considerations, ($\pm \varepsilon$)  and ($\pm 2\varepsilon$)  contribute equivalently, meaning that in practice, only three independent points are required for an accurate fitting procedure.

  \begin{table*}
\caption{  List of strains required for computing thermal expansions (T) and elastic constant tensors (E) for each crystallographic system. A star (*) next to a strain indicates that the strain is eliminated whenever symmetry permits.  Strains marked as E$^\diamond$ are specifically required for quadratic fitting of the vibrational free energy.  The strain magnitude is set as  $\delta \varepsilon=0.005$.}

        \begin{ruledtabular}
                \begin{tabular}{llccccccccccccccccccc}
Strain                      &     & Triclinic  & Monoclinic  &  Orthorhombic  &  Hexagonal &  Trigonal & Tetragonal &  Cubic \\
(0                         ,0                         ,0                         ,0                         ,0                         ,0)                           &  All   &  T E       &  T E        &  T E           &  T E       &  T E      &  T E       &  T E\\
( $\delta \varepsilon_{xx}$,0                         ,0                         ,0                         ,0                         ,0)                           &  $C_{11}$   &  T E       &  T E        &  T E           &  - E       &  - E      &  - E       &  - E\\
($-\delta \varepsilon_{xx}$,0                         ,0                         ,0                         ,0                         ,0)                           &  $C_{11}$   &  T E       &  T E        &  T E           &  - E       &  - E      &  - E       &  - E\\
($-\delta \varepsilon_{xx}$,$-\delta \varepsilon_{yy}$,0                         ,0                         ,0                         ,0)                           &  $C_{12}$   &  T E       &  T E        &  T E           &  T E       &  T E      &  T E       &  - E\\

($-\delta \varepsilon_{xx}$,0                         ,$-\delta \varepsilon_{zz}$,0                         ,0                         ,0)                           &  $C_{13}$   &  T E       &  T E        &  T E           &  - E       &  - E      &  - E       &  - -\\

($-\delta \varepsilon_{xx}$,0                         ,0                         ,$-\delta \varepsilon_{yz}$,0                         ,0)                           &  $C_{14}$   &  T E       &  - -        &  - -           &  - -       &  - -      &  - -       &  - -\\

($-\delta \varepsilon_{xx}$,0                         ,0                         ,0                         ,$-\delta \varepsilon_{xz}$,0)                           &  $C_{15}$   &  T E       &  T E        &  - -           &  - -       &  - -      &  - -       &  - -\\

($-\delta \varepsilon_{xx}$,0                         ,0                         ,0                         ,0                         ,$-\delta \varepsilon_{xy}$)  &  $C_{16}$   &  T E       &  - -        &  - -           &  - -       &  - -      &  - -       &  - -\\

(0                         ,$ \delta \varepsilon_{yy}$,0                         ,0                         ,0                         ,0)                           &  $C_{22}$   &  T E       &  T E        &  T E           &  - -       &  - -      &  - -       &  - -\\
(0                         ,$-\delta \varepsilon_{yy}$,0                         ,0                         ,0                         ,0)                           &  $C_{22}$   &  T E       &  T E        &  T E           &  - -       &  - -      &  - -       &  - -\\

(0                         ,$-\delta \varepsilon_{yy}$,$-\delta \varepsilon_{zz}$,0                         ,0                         ,0)                           &  $C_{23}$   &  T E       &  T E        &  T E           &  - -       &  - -      &  - -       &  - -\\

(0                         ,$-\delta \varepsilon_{yy}$,0                         ,$-\delta \varepsilon_{yz}$,0                         ,0)                           &  $C_{24}$   &  T E       &  - -        &  - -           &  - -       &  - -      &  - -       &  - -\\

(0                         ,$-\delta \varepsilon_{yy}$,0                         ,0                         ,$-\delta \varepsilon_{xz}$,0)                           &  $C_{25}$   &  T E       &  T E        &  - -           &  - -       &  - -      &  - -       &  - -\\

(0                         ,$-\delta \varepsilon_{yy}$,0                         ,0                         ,0                         ,$-\delta \varepsilon_{xy}$)  &  $C_{26}$   &  T E       &  - -        &  - -           &  - -       &  - -      &  - -       &  - -\\

(0                         ,0                         ,$ \delta \varepsilon_{zz}$,0                         ,0                         ,0)                           &  $C_{33}$   &  T E       &  T E        &  T E           &  T E       &  T E      &  T E       &  - -\\
(0                         ,0                         ,$-\delta \varepsilon_{zz}$,0                         ,0                         ,0)                           &  $C_{33}$   &  T E       &  T E        &  T E           &  T E       &  T E      &  T E       &  - -\\
(0                         ,0                         ,$-\delta \varepsilon_{zz}$,$-\delta \varepsilon_{yz}$,0                         ,0)                           &  $C_{34}$   &  T E       &  - -        &  - -           &  - -       &  - -      &  - -       &  - -\\

(0                         ,0                         ,$-\delta \varepsilon_{zz}$,0                         ,$-\delta \varepsilon_{xz}$,0)                           &  $C_{35}$   &  T E       &  T E        &  - -           &  - -       &  - -      &  - -       &  - -\\
(0                         ,0                         ,$-\delta \varepsilon_{zz}$,0                         ,0                         ,$-\delta \varepsilon_{xy}$)  &  $C_{36}$   &  T E       &  - -        &  - -           &  - -       &  - -      &  - -       &  - -\\
(0                         ,0                         ,0                         , $\delta \varepsilon_{yz}$,0                         ,0)                           &  $C_{44}$   &  T E       &  - -       &  - E           &  - E       &  - E      &  - E       &  - E\\
(0                         ,0                         ,0                         ,$-\delta \varepsilon_{yz}$,0                         ,0)                           &  $C_{44}$*  &  T E       &  - E        &  - -           &  - -       &  - -      &  - -       &  - -\\
(0                         ,0                         ,0                         ,$-\delta \varepsilon_{yz}$,$-\delta \varepsilon_{xz}$,0)                           &  $C_{45}$   &  T E       &  - -        &  - -           &  - -       &  - -      &  - -       &  - -\\
(0                         ,0                         ,0                         ,$-\delta \varepsilon_{yz}$,0                         ,$-\delta \varepsilon_{xy}$)  &  $C_{46}$   &  T E       &  - E       &  - -           &  - -       &  - -      &  - -       &  - -\\

(0                         ,0                         ,0                         ,0                         , $\delta \varepsilon_{xz}$,0)                           &  $C_{55}$   &  T E       &  T E        &  - E           &  - -       &  - -      &  - -       &  - -\\
(0                         ,0                         ,0                         ,0                         ,$-\delta \varepsilon_{xz}$,0)                           &  $C_{55}$*  &  T E       &  T E        &  - -           &  - -       &  - -      &  - -       &  - -\\
(0                         ,0                         ,0                         ,0                         ,$-\delta \varepsilon_{xz}$,$-\delta \varepsilon_{xy}$)  &  $C_{56}$   &  T E       &  - -        &  - -           &  - -       &  - -      &  - -       &  - -\\
(0                         ,0                         ,0                         ,0                         ,0                         ,$ \delta \varepsilon_{xy}$)  &  $C_{66}$   &  T E       &  - -        &  - E           &  - -       &  - -      &  - E       &  - -\\
(0                         ,0                         ,0                         ,0                         ,0                         ,$-\delta \varepsilon_{xy}$)  &  $C_{66}$*  &  T E       &  - E        &  - -           &  - -       &  - -      &  - -       &  - -\\
( $\delta \varepsilon_{xx}$,$ \delta \varepsilon_{yy}$,0                         ,0                         ,0                         ,0)                           &  - -   &  - -       &  - -        &  - -           &  T -       &  T -      &  T -       &  - -\\
( $\delta \varepsilon_{xx}$,$ \delta \varepsilon_{yy}$,$ \delta \varepsilon_{zz}$,0                         ,0                         ,0)                           &  - -   &  - -       &  - -        &  - -           &  - -       &  - -      &  - -       &  T -\\
($-\delta \varepsilon_{xx}$,$-\delta \varepsilon_{yy}$,$-\delta \varepsilon_{zz}$,0                         ,0                         ,0)                           &  - -   &  - -       &  - -        &  - -           &  T -       &  T -      &  T -       &  T -\\
(0                         ,0                         ,0                         , $2\delta \varepsilon_{yz}$,0                         ,0)                           &  $C_{44}$$^\diamond$   &  - -      &  - -       &  - E           &  - E       &  - E      &  - E       &  - E\\
(0                         ,0                         ,0                         ,0                         , $2\delta \varepsilon_{xz}$,0)                           &  $C_{55}$$^\diamond$   &  - -       &  - -       &  - E           &  - -       &  - -      &  - -       &  - -\\
(0                         ,0                         ,0                         ,0                         ,0                         ,$2 \delta \varepsilon_{xy}$)  &  $C_{66}$$^\diamond$   &  - -      &  - -        &  - E           &  - -       &  - -      &  - E       &  - -\\
($-\delta \varepsilon_{xx}$,0                         ,0                         ,$\delta \varepsilon_{yz}$,0                         ,0)                           &  $C_{14}$$^\diamond$   &  - -       &  - -        &  - -           &  - -       &  - E      &  - -       &  - -\\

    \end{tabular}
        \end{ruledtabular}\label{tab:strain}
\end{table*}

\bibliography{ThermalExpansion2}

\end{document}